\documentclass[10pt,journal,compsoc]{IEEEtran}
\usepackage{algorithm}
\usepackage{algpseudocode}
\usepackage{algorithmicx}

\usepackage{booktabs}
\usepackage{graphicx}
\graphicspath{{./figure}}
\DeclareGraphicsExtensions{.pdf}
\usepackage{subfigure}
\usepackage{epstopdf}
\usepackage{amsmath}
\usepackage{cite}
\usepackage{mathtools} 
\usepackage{makecell}  
\usepackage{multirow}
\usepackage{verbatim}  
\usepackage[table]{xcolor}
\usepackage{colortbl}
\usepackage{array}
\usepackage{listings}

\usepackage{enumitem} 
\usepackage{enumerate}
\usepackage{soul}
\soulregister{\cite}7 
\soulregister{\citep}7 
\soulregister{\citet}7 
\soulregister{\ref}7 
\soulregister{\pageref}7 

\usepackage{pifont}
\usepackage{tikz}
\newcommand*{\circled}[1]{\lower.7ex\hbox{\tikz\draw (0pt, 0pt)%
		circle (.5em) node {\makebox[1em][c]{\small #1}};}}

\ifCLASSOPTIONcompsoc
 \usepackage[caption=false,font=footnotesize,labelfont=sf,textfont=sf]{subfig}
\else
 \usepackage[caption=false,font=footnotesize]{subfig}
\fi
\usepackage{stfloats}
\ifCLASSOPTIONcaptionsoff
 \usepackage[nomarkers]{endfloat}
\let\MYoriglatexcaption\caption
\renewcommand{\caption}[2][\relax]{\MYoriglatexcaption[#2]{#2}}
\fi
\usepackage{url}

\definecolor{ColorSaddleBrown}{RGB}{139,69,19} 
\definecolor{ColorDarkGray}{RGB}{169,169,169} 
\begin{document}

\title{\LARGE Dynamic GPU Energy Optimization for Machine Learning Training Workloads}

\author{
Farui~Wang, Weizhe~Zhang, Shichao~Lai, Meng~Hao, and~Zheng~Wang

\IEEEcompsocitemizethanks{
\IEEEcompsocthanksitem F. Wang, W. Zhang, S. Lai, and M. Hao are with the School of
Cyberspace Science at Harbin Institute of Technology, Harbin 150000, China.\protect\\
E-mail:\{wangfarui,wzzhang\}@hit.edu.cn
\IEEEcompsocthanksitem Z. Wang is with the School of Computing at University of Leeds, United Kingdom.\protect\\
E-mail: z.wang5@leeds.ac.uk}
}

\IEEEtitleabstractindextext{
\begin{abstract}
GPUs are widely used to accelerate the training of machine learning workloads. As modern machine learning models
become increasingly larger, they require a longer time to train, leading to higher GPU energy consumption. This paper
presents GPOEO, an online GPU energy optimization framework for machine learning training workloads. GPOEO
dynamically determines the optimal energy configuration by employing novel techniques for online measurement,
multi-objective prediction modeling, and search optimization. To characterize the target workload behavior, GPOEO
utilizes GPU performance counters. To reduce the performance counter profiling overhead, it uses an analytical model
to detect the training iteration change and only collects performance counter data when an iteration shift is
detected. GPOEO employs multi-objective models based on gradient boosting and a local search algorithm to find a
trade-off between execution time and energy consumption. We evaluate the GPOEO by applying it to 71 machine learning
workloads from two AI benchmark suites running on an NVIDIA RTX3080Ti GPU. Compared with the NVIDIA default
scheduling strategy, GPOEO delivers a mean energy saving of 16.2\% with a modest average execution time increase of
5.1\%.
\end{abstract}


}

\maketitle
\IEEEdisplaynontitleabstractindextext
\IEEEpeerreviewmaketitle

\IEEEraisesectionheading{\section{Introduction}}
In recent years, deep neural networks (DNNs) have demonstrated breakthrough effectiveness in various tasks that were once deemed impossible \cite{lecun2015deep,alom2019state}. Training an effective DNN requires performing expensive training on a large number of samples. As the model training time increases, the energy consumption of machine
learning training workloads also grows. This is a particular concern for high-performance GPUs because they are widely used to train DNN
workloads but are less energy-efficient than their CPU counterparts. As a result, there is a critical need to reduce the energy
consumption for GPUs when training machine learning workloads. Achieving this goal can reduce the maintenance cost and allow users to train
larger models within the same budget.

Most of the existing studies in general GPU energy optimization are  offline techniques
\cite{guerreiro2018modeling,guerreiro2019modeling,guerreiro2019classification,guerreiro2019ptx,arafa2020verified}. These techniques require
profiling applications ahead of time \cite{guerreiro2018modeling,guerreiro2019modeling,guerreiro2019classification,arafa2020verified} or
analyzing their source code \cite{guerreiro2019ptx} in advance to collect application features. These approaches cannot adapt to the change
of the program behavior and can incur expensive profiling overhead. Other approaches change the internal working mechanisms of machine
learning (ML) models \cite{wang2020energy,nabavinejad2021batchsizer}, but they are specific to the ML algorithms used and do not generalize
to other ML methods.

Recently, efforts have been made to target online GPU power optimization. The work presented in \cite{majumdar2017dynamic} proposes a
dynamic power management method for integrated APUs. ODPP \cite{ODPP} is another online dynamic power-performance management framework.
ODPP detects the periods of applications to build prediction models. However, their models can only detect coarse-grained phase change
patterns.

This work aims to provide a better dynamic GPU energy optimization method for iterative ML training workloads. We present GPOEO, a
micro-intrusive \footnote{Micro-intrusive means we require minor changes to the program source code.} GPU online energy optimization
framework. GPOEO only requires the user to mark the beginning and the end of the target code region. GPOEO can then automatically collect
the relevant information to monitor the program behavior to identify the repeat workload patterns and adjust the energy optimization scheme
accordingly.

GPOEO trades execution time for energy efficiency. To this end, we employ multi-objective optimization models built on the XGBoost
Classifier \cite{chen2016xgboost}. As a departure from existing online GPU power optimization schemes, GPOEO leverages the fine-grained
runtime information provided by the hardware performance counters to model the system and program behavior to build more accurate decision
models. We use the predictive model as a cost function to search for the optimal energy configuration for the current workload pattern.
These supporting methods enable us to develop a period detection algorithm based on the fast Fourier transform (FFT) and feature sequence
similarity evaluation to identify if a changed training iteration takes place and adjust our optimization accordingly.

We evaluate GPOEO by applying it to 71 machine learning workloads from three AI benchmark suites \cite{tang2021aibench}. We test our
approach on an NVIDIA RTX3080Ti GPU. Experimental results show that GPOEO can reduce GPU energy consumption by over 20\%
(16.2\% on average), with most execution time increments less than 7\% (5.1\% on average) when compared to the NVIDIA default
GPU scheduling strategy.

This paper makes the following contributions:

\begin{itemize}
\item It is the first work to exploit performance counter metrics to perform online energy prediction and optimization for discrete GPUs.
By minimizing the instrument and measurement overhead, our approach improves the practicability of online GPU profiling.

\item It presents a robust detection algorithm to automatically detect the iterative period of ML applications from
    the GPU resource traces (Section \ref{ssec:period_detection}).

\item It showcases how an effective GPU online energy optimization framework can be constructed as a cost model to quickly identify a
good energy configuration.

\end{itemize}

\noindent \textbf{Online material:} The GPOEO framework is publicly available at \url{https://github.com/ruixueqingyang/GPOEO}.

\section{Background and Motivation}

\subsection{GPU configuration and profiling tools} \label{ssec:tools}

NVIDIA provides the NVML \cite{NVML} and CUPTI \cite{CUPTI} library for reconfiguring, measuring, and profiling its GPUs. The NVML
\cite{NVML} can set Streaming Multiprocessor (SM) clock frequency, global memory clock frequency (only certain types of NVIDIA GPUs), and
powercap online. It also supports measuring power, SM utilization, and global memory utilization. The CUPTI \cite{CUPTI} can profile
performance counter. These tools make GPU online energy optimization possible. In this work, we adjust the clock frequency of the GPU
SM and global memory to achieve energy saving.


\subsection{Motivation}

\subsubsection{Energy Optimization Potential}
\begin{figure}[t]
	\centering
	\includegraphics[width=3.5in]{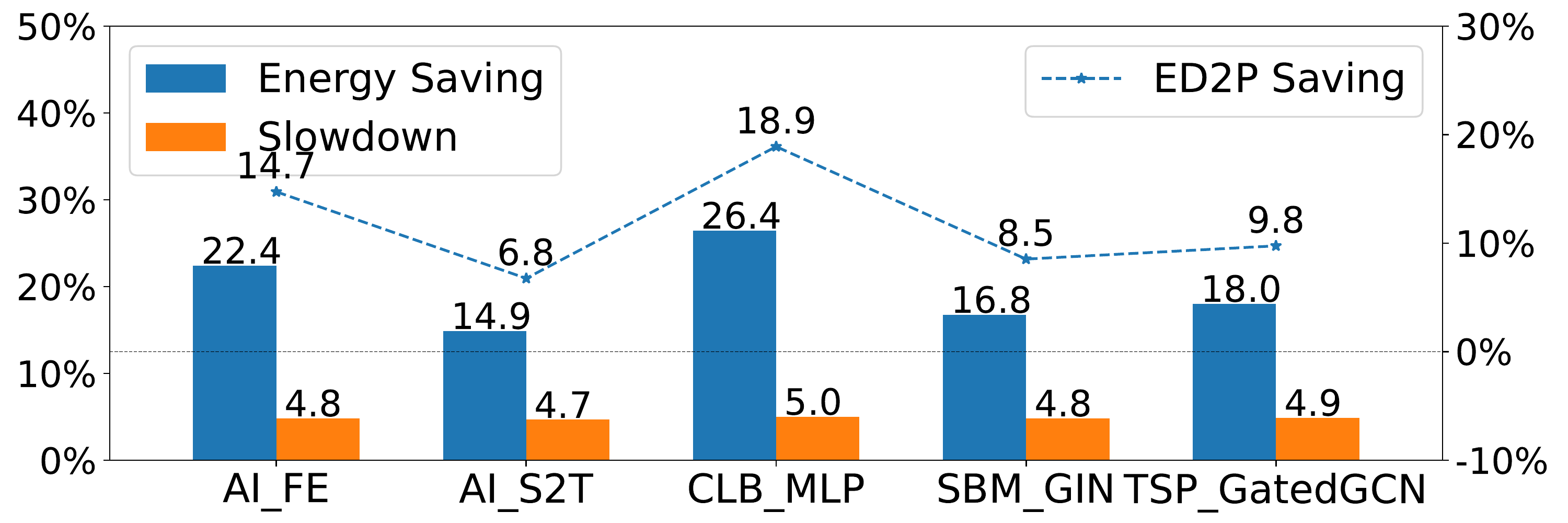}
	\caption{Oracle ED2P saving of ML applications}
	\label{fig:potential}
\end{figure}

The training phase of machine learning usually uses high-performance GPUs as accelerators. GPUs consume a lot of energy and time. To
explore the energy-saving potential on GPUs, we run five applications in AIBench \cite{tang2021aibench} and benchmarking-gnns
\cite{dwivedi2020benchmarking} on all combinations of SM and memory clock frequencies and select the best configurations for minimal
energy consumption within the slowdown constraint of 5\%. Figure \ref{fig:potential} shows the oracle results of energy saving, slowdown, and
ED2P ($Energy \times Dealy^{2}$) saving. The AI\_FE, AI\_S2T, and SBM\_GIN are relatively compute-intensive and save significant energy
(14.9\%-22.4\%). The CLB\_MLP and TSP\_GatedGCN are relatively memory-intensive and save considerable energy (18.0\%-26.4\%). These results
highlight that both compute-intensive and memory-intensive ML training workloads have the chance to save energy with acceptable slowdown.

\subsubsection{Make Energy Optimization Practical}
Most existing studies on GPU energy optimization are offline and need domain knowledge and complex offline profiling. Thus, users or
researchers can hardly apply these studies in practice. An ideal online GPU energy optimization system should be non-intrusive to
applications and transparent to users.
Majumdar's work \cite{majumdar2017dynamic} only applies to AMD APUs. 
Motivated by drawbacks, we propose an online energy optimization framework called GPOEO. Users only need to insert the Begin and End APIs,
and the GPOEO automatically optimizes energy online.


\begin{figure} \centering
	\subfigure[MLC\_3WLGNN application] {
		\includegraphics[width=3.5in]{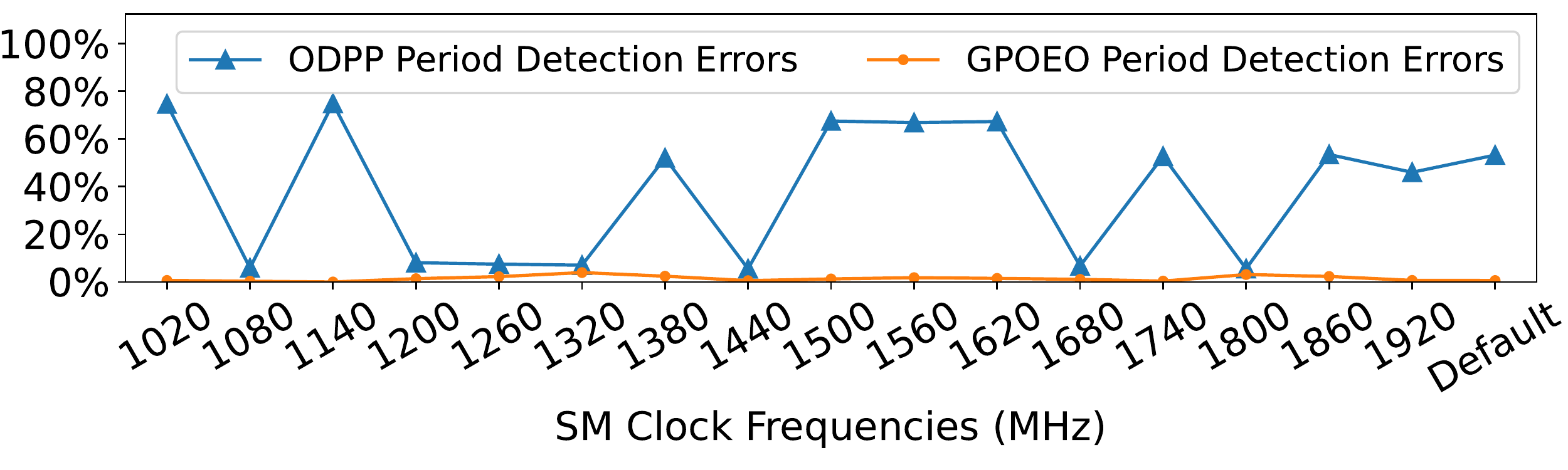}
	}
	\subfigure[SP\_GCN application] {
		\includegraphics[width=3.5in]{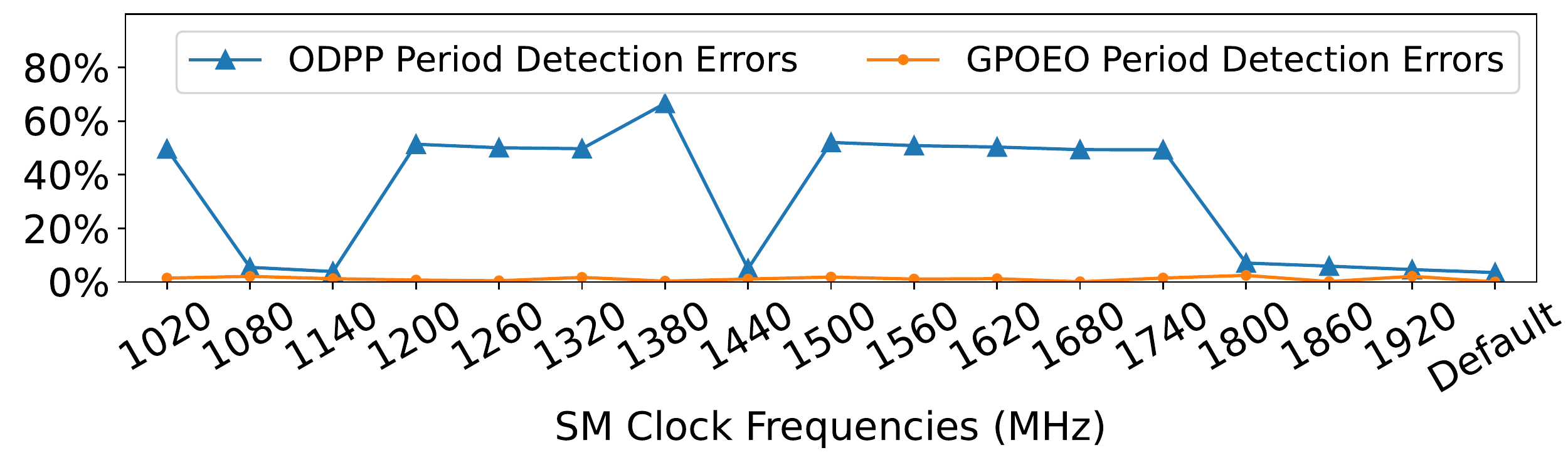}
	}
	\caption{Period detection errors of ODPP and GPOEO under different SM clock frequencies}
	\label{fig:Period_All_Clocks}
\end{figure}

%

\subsubsection{Reduce Period Detection Error} \label{sssec:fft_error}

Online sampling time series of power and utilization can reflect the periodic pattern. ODPP \cite{ODPP} uses the Fourier transform
algorithm to detect periods, namely the execution time of one major iteration. However, according to our experimental results, the period
error of ODPP \cite{ODPP} is quite large when the periodicity is not apparent. To solve this shortcoming, we propose a robust period
detection algorithm. Figure \ref{fig:Period_All_Clocks} shows the absolute percentage
errors of period detection with ODPP and our algorithm GPOEO under different SM clock frequencies on two ML applications. The ``Default" means the NVIDIA
default scheduling strategy. The period detection errors of ODPP are pretty significant and unstable, while the period detection errors of
our GPOEO are less than 5\% under all SM clock frequencies.

\subsubsection{Using GPU Performance Counters} \label{sssec:counter}
\begin{figure}[t]
	\centering
	\includegraphics[width=3.5in]{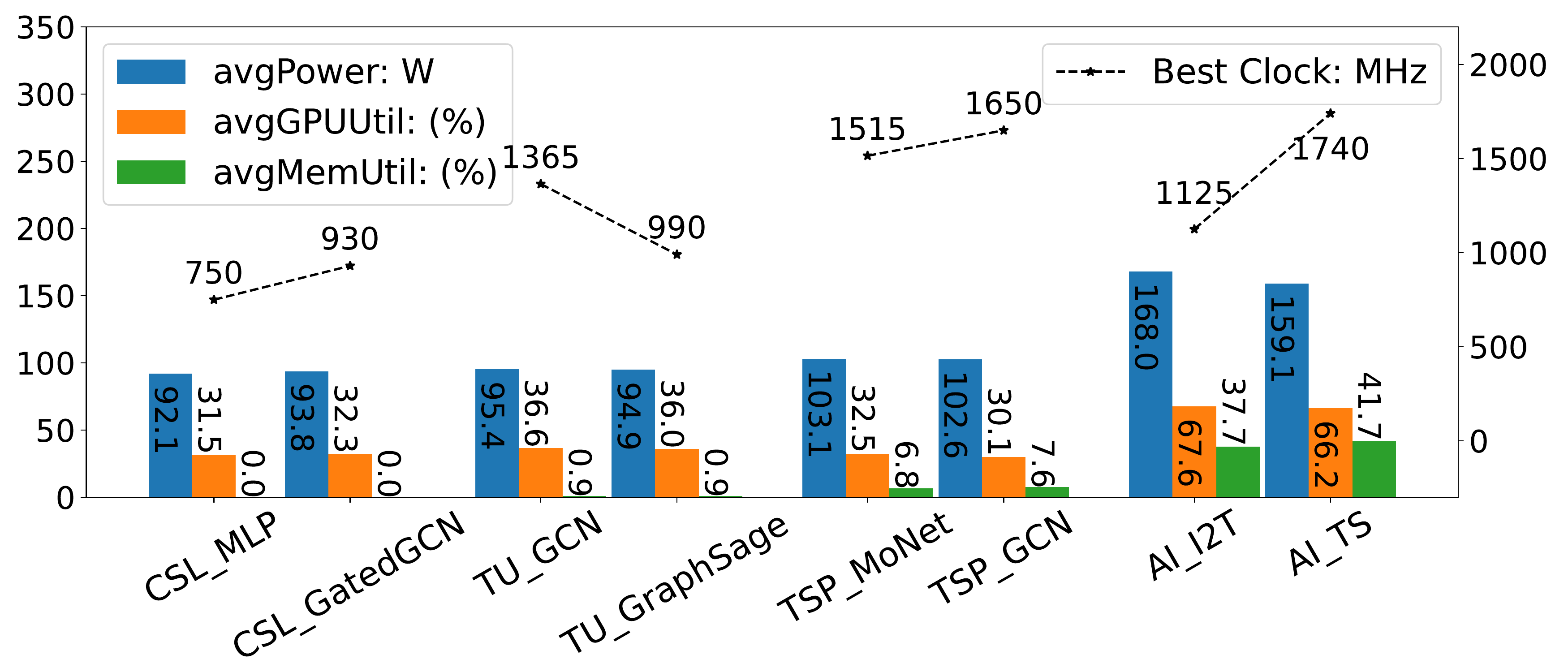}
	\caption{ML applications with similar coarse-grained features and different optimal SM clocks for ED2P}
	\label{fig:toplevelfeature}
\end{figure}

ODPP \cite{ODPP} uses power, SM utilization, and memory utilization as features to predict energy and execution time.
However, these coarse-grained features are inadequate to give accurate predictions for some machine learning applications. We give an
example in Figure \ref{fig:toplevelfeature}. Applications in each pair have similar average power and utilization under the same reference
clock frequency configuration, but their optimal SM clocks for ED2P are different (memory frequency fixed at 9251MHz). In light of this observation, we wish to use more
fine-grained metrics to capture the subtle interactions between the application and the hardware. To this end, we use hardware performance
counter metrics as features to build predictive models to estimate energy and execution time. Later in the paper, we show that this
strategy allows us to build more accurate prediction models, which in turn lead to better optimization decisions. The key challenges here
are reducing the overhead of performance counter profiling and minimizing code modifications.

\section{Overview of Our Approach}
This section defines the scope of the work and gives an overview of our approach.
\subsection{Problem Formulation}

This work aims to find an optimal SM and memory clock frequency configuration to balance the GPU energy consumption and execution time for
ML training workloads. Our goal can be formulated as follows:

\begin{equation} \label{equ:SM_formalize}
\begin{split}
\mathrm{min} \quad F_{SM} &= f_{obj}\left ( \widehat{eng_{SM}},\widehat{time_{SM}} \right ) \\
\mathrm{s.t.} \ \widehat{eng_{SM}} &= EngMdl_{SM}\left ( \textbf{w}_{SM} \right ), \\
\widehat{time_{SM}} &= TimeMdl_{SM}\left ( \textbf{w}_{SM} \right ), \\
\textbf{w}_{SM} &= \left \{ SMgear_i, \textbf{Feature} \right \}, \\
SMgear_i &\in \left \{ SMgear_{1},\dots,SMgear_{m} \right \}, \\
\textbf{Feature} &= \left \{ feature_{1},\dots,feature_{q} \right \}.
\end{split}
\end{equation}

\begin{equation} \label{equ:Mem_formalize}
	\begin{split}
		\mathrm{min} \quad F_{Mem} &= f_{obj}\left ( \widehat{eng_{Mem}},\widehat{time_{Mem}} \right ) \\
		\mathrm{s.t.} \ \widehat{eng_{Mem}} &= EngMdl_{Mem}\left ( \textbf{w}_{Mem} \right ), \\
		\widehat{time_{Mem}} &= TimeMdl_{Mem}\left ( \textbf{w}_{Mem} \right ), \\
		\textbf{w}_{Mem} &= \left \{ Memgear_i, \textbf{Feature} \right \}, \\
		Memgear_i &\in \left \{ Memgear_{1},\dots,Memgear_{p} \right \}.
	\end{split}
\end{equation}

Table \ref{tab:notation} lists the variables used in our problem formulation. We note that our approach can be applied
to an arbitrary objective function. We select the energy consumption with an execution time increase constraint as our
objective function to explicitly control the performance loss. Like \cite{schwarzrock2020runtime}, we assume that the
search space of SM and memory clocks is convex and optimize the SM frequency and memory frequency in order. We use four
prediction models to estimate energy and time with different SM and memory clock frequencies. Then we select the SM and
memory clock configuration that can best satisfy the optimization goal.

\begin{table}
	\footnotesize
	\caption{Description of Parameters}
	\label{tab:notation}
	\begin{center}
\rowcolors{2}{gray!25}{white}
	\begin{tabular}{lp{0.7\columnwidth}}
		\toprule
		Notation & Description                                                   \\
        \midrule
		$f_{obj}$     & objective function                                            \\
		$\widehat{eng_{SM}}$      & energy consumption predicted with the SM clock     \\
		$\widehat{time_{SM}}$     & execution time predicted with the SM clock         \\
		$EngMdl_{SM}$     & energy prediction model with the SM clock        \\
		$TimeMdl_{SM}$     & time prediction model with the SM clock         \\
		$SMgear_i$     & one SM gear represents an SM clock frequency                    \\
		$\textbf{w}_{SM}$        & the input vector containing SM clock        \\
		
		$\widehat{eng_{Mem}}$      & energy consumption predicted with the memory clock          \\
		$\widehat{time_{Mem}}$     & execution time predicted with the memory clock              \\
		$EngMdl_{Mem}$     & energy prediction model with the memory clock   \\
		$TimeMdl_{Mem}$     & time prediction model with the memory clock    \\
		$Memgear_i$     & one memory gear represents a memory clock frequency                   \\
		$\textbf{w}_{Mem}$        & the input vector containing memory clock                         \\
		
		$\textbf{Feature}$  & feature vector measured under the reference clock configuration from the ML applications \\
        \bottomrule
	\end{tabular}
	\end{center}
\end{table}

\begin{figure}[!t]
	\centering
	\includegraphics[width=3.5in]{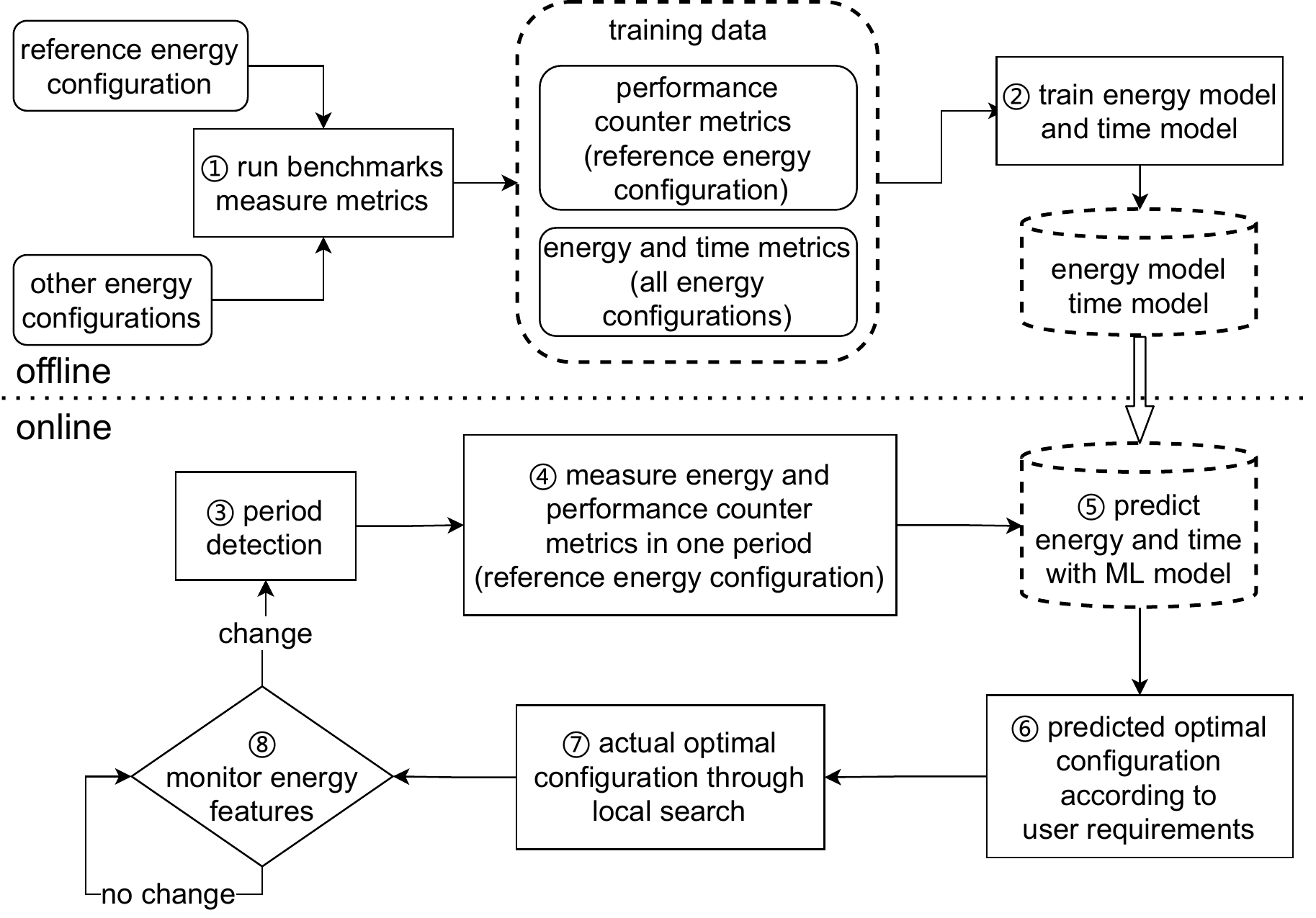}
	\caption{Overview of the GPOEO workflow.}
	\label{fig:overview}
\end{figure}

\subsection{Framework Overview}
GPOEO includes two stages, shown in Figure \ref{fig:overview}. In the offline training stage, we run representative benchmarks on each SM
and memory clock frequencies (\circled{1}) to collect performance counter metrics on the reference SM and memory frequencies and
energy-time data on all frequencies. Then we train multi-objective models (\circled{2}). In the online optimization stage, we measure energy and performance counter metrics in one
detected period (\circled{3}, \circled{4}), then predict the optimal SM and memory frequencies (\circled{5}, \circled{6}). After that, we
try frequencies around the predicted optimal configuration and compare the measured energy-time data to search for the actual optimal
configuration (\circled{7}). Finally, we set the actual optimal configuration and monitor energy characteristics (\circled{8}). If the
fluctuation of energy characteristics exceeds the threshold, another optimization process will start.

\section{Design and Implement}

\subsection{Robust Period Detection} \label{ssec:period_detection}
Iterative ML workloads exhibit periodic behavior across iterations.
We treat one period, namely one iteration, as basic unit for online metric measurement and evaluation. Our approach
does not rely on offline code instrumentation to reduce developer involvement.

\subsubsection{Fourier Transform: Get Candidate Periods} \label{sssec:fft_candidate}
To detect the period change, we leverage the Fourier transform. This established methodology is widely used in signal
processing to detect the dominant frequency and period. In our scenario, we assume the periodic feature of ML
applications, such as GPU power and utilization, can be expressed as a time-domain function $\mathcal{F}(t)$. Generally
speaking, $\mathcal{F}(t)$ is periodic and can be expressed as a linear combination of trigonometric series (know as
Fourier Series) in exponential form as:
\begin{equation} \nonumber
\mathcal{F}\left( t \right) = \sum_{n = -\infty}^{\infty}C_{n}e^{{2\pi jnt}/{T_{iter}}}
\end{equation}
where $C_{n}$ are Fourier coefficients.

We can get a series of frequency components via Fourier transform:
\begin{equation} \nonumber
\mathcal{F}(\omega) = 2\pi \sum_{n=-\infty}^{\infty}C_{n}\delta\left(\omega - 2\pi n/T_{iter}\right)
\end{equation}
the frequency $\mathcal{F}(\omega)$ is a series of impulse functions (i.e., $\delta$) located at $2\pi n/T_{iter}$, with amplitude proportional to $C_{n}$. These impulses represent different frequency components in $\mathcal{F}(t)$. The one with the largest amplitude is the major frequency component. The iterative period $T_{iter}$ of the ML application can be calculated as $T_{iter} = 1/f_{major}$, where $f_{major}$ is the frequency of the major frequency component.

\subsubsection{Feature Sequence Similarity: Find Accurate Periods} \label{sssec:sequence_similarity}

As mentioned in Section \ref{sssec:fft_error}, the period detection results of the Fourier transform may have vast errors.
%
%
Our period calculation algorithm (Algorithm \ref{alg:Calcu}) combines the fast Fourier transform (FFT) and feature sequence similarity (Algorithm \ref{alg:similarity}) to reduce these errors.
We find amplitude peaks and corresponding periods $TPeak$ in the result of FFT (line 1-3). Then we select periods $Tpeak_m$ with relatively high peaks as candidate periods $Tcand$ (line 4-5). We set the peak coefficient $c_{peak}$ to 0.6-0.7 empirically.
Then we use feature sequence similarity (Algorithm \ref{alg:similarity}) to evaluate and select the best candidate period with minimal error.
Finally, we perform a local search around the best candidate period to further improve accuracy.

\begin{algorithm}
	\caption{Period calculation algorithm based on FFT and feature sequence similarity}
	\label{alg:Calcu}
	{\bf Input:} feature sampling sequence $\textbf{Smp} = \{s_1, \dots, s_N\}$, sampling interval $T_s$\\
	{\bf Output:} iterative period $T_{iter}$, error of period $err$\\
	\begin{algorithmic}[1]
		\State $Freq, Ampl = \{freq_i\}, \{ampl_i\}=$FFT$(\textbf{Smp}, T_s)$
		\State $T=\{T_i\}=\{1/freq_i\}$
		\State Find peaks in $Ampl$, get the set of peaks $AmplPeak=\{amplpeak_m\}$ and the set of corresponding periods $TPeak = \{Tpeak_m\}$
		\State $IndexCand = \{idxcand_k\}$
		\Statex $\qquad \qquad = \{m|amplpeak_m > c_{peak} \times max(AmplPeak)\}$
		\State $TCand = \{Tcand_p\} = \{Tpeak_{idxcand_k}\}$
		\For{$Tcand_p \in TCand$}
		\State $err_p =$ Algorithm2($Tcand_p, \textbf{Smp}, T_s$)
		\EndFor
		\State Find $err_{idxmin} = min(\{err_p\})$
		\State $Tcand_{opt} = Tcand_{idxmin}$
		\State $N_T = T_s \times (N-1)/Tcand_{opt},$
		\Statex $T_{low} = Tcand_{opt}\times (1-1/(N_T + 1))$
		\State $T_{up} = Tcand_{opt}\times (1 + 1/(N_T - 1))$
		\State Construct an arithmetic sequence $TLocal = \{Tlocal_q\},$
		\Statex where $Tlocal_q = T_{low} + (q-1)\times T_{accu},$
		\Statex $\qquad \quad q = 1, \dots, (T_{up}-T_{up})/T_{accu}$
		\For{$Tlocal_q \in TLocal$}
		\State $errlocal_q=$Algorithm2($Tlocal_p, \textbf{Smp}, T_s$)
		\EndFor
		\State Find $err = err_{idxmin} = min(\{errlocal_q\})$
		\State $T_{iter} = Tlocal_{idxmin}$ \\
		\Return $T_{iter}, err$
	\end{algorithmic}
\end{algorithm}

We propose the feature sequence similarity algorithm (Algorithm \ref{alg:similarity}) to evaluate candidate periods.
We divide the feature sampling curve into several sub-curves, and the time duration of each sub-curve is equal to the candidate period (line 1-4). The more similar the curves are, the closer the candidate period is to the actual period.
So we evaluate the similarity of each pair of adjacent sub-curves (line 5-18).

Each sub-curve is a time series, and calculating the Euclidean distance between corresponding points is a common method to
measure the similarity of time series. However, in our scenario, the Euclidean distance is susceptible to high-frequency
interference and reports the wrong similarity.
Inspired by this, instead of calculating the distance of each pair of corresponding points, we group the sampling data within a
series and evaluate the distance between the corresponding groups. Specifically, we use the Gaussian mixture model algorithm to
cluster each sub-curves into N groups (line 8-11). Then, for each group, we calculate the relative average amplitudes (line
12-13). The average operation can eliminate the influence of high-frequency interference. Later, we calculate the symmetric mean
absolute percentage error (SMAPE) of the current group pair (line 14). Finally, we calculate the weighted average of all SMAPEs
as similarity error (line 16-17). The weights are the numbers of sampling points in groups. When all adjacent sub-curves are
evaluated, we treat the mean similarity error as the error of $T_iter$ (line 19).

\begin{algorithm}
	\caption{Feature sequence similarity algorithm}
	\label{alg:similarity}
	{\bf Input:} Period to be evaluated $T_{iter}$, feature sampling sequence $\textbf{Smp} = \{s_1, \dots, s_N\}$, sampling interval $T_s$\\
	{\bf Output:} error of period $Err_T$
	\begin{algorithmic}[1]
		\State $Num_T = \lfloor s_N/T_{iter}\rfloor; Num_s = \lfloor T/T_s \rfloor$
		\For{$i \in \{1, \dots Num_T\}$}
		\State $\textbf{Smp}_i = \{s_{(i-1)\times Num_s+1}, \dots, s_{i \times Num_s}\} = \{ss_{i,j}\}$
		\Statex \qquad \qquad \qquad \qquad \qquad \qquad \qquad \qquad $(j=1,\dots, Num_s)$
		\EndFor
		\For{$i \in \{1, \dots, Num_T - 1\}$}
		\State $Mean_{prev} = mean(\textbf{Smp}_i)$
		\State $Mean_{back} = mean(\textbf{Smp}_{i+1})$
		\State $\{\textbf{GaGrp}_1, \dots, \textbf{GaGrp}_{NumG}\} = $Gauss$(\textbf{Smp}_i, NumG)$
		\Statex \qquad \qquad \qquad \, where $\textbf{GaGrp}_j = \{idx_{j,1}, \dots, idx_{j, Numj}\}$
		\For{$j \in \{1, \dots, NumG\}$}
		\State $\textbf{Grp}_{prev} = \{ss_{i, idx_{j,1}}, \dots, ss_{i, idx_{j, Numj}}\}$
		\State $\textbf{Grp}_{back} = \{ss_{i+1, idx_{j,1}}, \dots, ss_{i+1, idx_{j, Numj}}\}$
		\State $RelValPrev_j = mean(\textbf{Grp}_{prev}) - Mean_{prev}$
		\State $RelValBack_j = mean(\textbf{Grp}_{back}) - Mean_{back}$
		\State $grperr_j = $SMAPE$(RelValPrev_j, RelValBack_j)$
		\EndFor
		\State $Weight = \{|GaGrp_1|, \dots,|GaGrp_{NumG}|\}$
		\State $err_i = avg(\{grperr_1, \dots, grperr_{NumG}\}, Weight)$
		\EndFor\\
		\Return $Err_T = mean(\{err_1, \dots, err_{Num_T-1}\})$
	\end{algorithmic}
	\emph{Note: Gauss = the Gaussian mixture model clustering algorithm}
\end{algorithm}

\subsubsection{Online Robust Period Detection Algorithm Framework}
To get accurate $T_{iter}$ dynamically, we design an online robust period detection algorithm framework (Algorithm \ref{alg:detection}).
The framework calculates $T_{iter}$ in a rolling form while sampling the power and utilization of the GPU until the $T_{iter}$ is
stable. First, we get the initial period $T_{init}$, using Algorithm \ref{alg:similarity}.
If sampling duration is short than $c_{measure}$ times $T_{init}$ ($c_{measure}$ is set to 2 empirically), it is too short to do a rolling
calculation, so we calculate the required sampling duration, skip rolling calculation, and return (line2-6). If the sampling duration is long
enough, we get some updated samples by setting $t_{start}$ (line 7). Samples before $t_{start}$ are ignored because they may be outdated. We
set $step=0.5$ and $c_{eval}=6.5$ according to our experiments.
Then, the rolling period calculation begins. In each iteration, we calculate the period (line 9-11) and increase $t_{start}$
to exclude more outdated samples for the next iteration (line 12). With the rolling period calculation, we get a sequence of periods with
errors (line 14). If an ML application runs in the regular iteration phase, these periods should be similar.
If the difference of these periods is less than the threshold, set $SmpDur_{next}=-1$ to indicate stop sampling. Otherwise, calculate $SmpDur_{next}$ (line 16-21). The best period, namely $T_{iter}$, is the period with the minimal error (line 15).
If $SmpDur_{next}>0$, we ignore $T_{iter}$, keep sampling, and call Algorithm \ref{alg:detection} again after $SmpDur_{next}$. Otherwise, we use $T_{iter}$ for feature measurement (Section \ref{ssec:measurement}).

\begin{algorithm}
	\caption{Online robust period detection algorithm framework}
	\label{alg:detection}
	{\bf Input:}feature sampling sequence $\textbf{Smp} = \{s_1, \dots, s_N\}$, sampling interval $T_s$\\
	{\bf Output:}iterative period $T_{iter}$, next sampling duration $SmpDur_{next}$
	\begin{algorithmic}[1]
		\State $T_{init}, err_{init} =$Algorithm1($\textbf{Smp}, T_s$)
		\State $SmpDur=(N-1)\times T_s$
		\If{$SmpDur < c_{measure} \times T_{init}$}
		\State $T_{iter} = T_{init}$
		\State $SmpDur_{next} = c_{measure} \times T_{init} - SmpDur$
		\EndIf
		\State $t_{start} = max(0, (SmpDur - (2 + c_{eval}\times step)\times T_{init}))$
		\While{$(SmpDur - t_{start}) / T_{init}\ge c_{measure}$}
		\State $istart = 1 + \lfloor t_{start}/T_s\rfloor$
		\State $\textbf{SubSmp} = \{s_{istart}, \dots, s_N\}$
		\State $T_j, err_j = $Algorithm1($\textbf{SubSmp}, T_s$)
		\State $t_{start} = t_{start} + step \times T_{init}$
		\EndWhile
		\State $T=\{T_j\}, Err = \{err_j\}$
		\State Find $min(Err) = err_k$ then $T_{iter} = T_k$
		\State $Diff = abs((max(T) - min(T)) / mean(T))$
		\If{$Diff < Diff_{threshold}$}
		\State $SmpDur_{next} = -1$
		\Else
		\State $SmpDur_{next} = $
		\Statex $\qquad \lceil SmpDur / max(T) \rceil \times max(T) - SmpDur$
		\EndIf \\
		\Return $T_{iter}, SmpDur_{next}$
	\end{algorithmic}
\end{algorithm}

\begin{algorithm}
	\caption{Adaptive feature measurement}
	\label{alg:measurement}
	{\bf Input:} features to be measured $FeatureName = \{name_1, \dots, name_n\}$, the feature for iterative period detection $\textbf{Feature}_{dect}$\\
	{\bf Output:} feature data $\textbf{Feature} = \{feature_1, \dots, feature_n\}$
	\begin{algorithmic}[1]
		\State Start measurement for $FeatureName$ and $\textbf{Feature}_{dect}$
		\State $SmpDur_{next} = SmpDur_{init}$
		\While{$SmpDur_{next} > 0$}
		\State Delay $SmpDur_{next}$
		\State Collect sampling sequence of $\textbf{Feature}_{dect}$
		\State $T_{iter}, SmpDur_{next}=$Algorithm3$(\textbf{Feature}_{dect}, T_s)$
		\EndWhile
		\State Restart measurement for $FeatureName$
		\State Delay $T_{iter}$ then stop measurement
		\State Collect feature data $\textbf{Feature}$\\
		\Return $\textbf{Feature}$
	\end{algorithmic}
\end{algorithm}

\subsection{Micro-intrusive Online Feature Measurement} \label{ssec:measurement}
Based on the robust period detection, we propose the adaptive feature measurement algorithm (Algorithm \ref{alg:measurement}). The feature
measurement causes overhead and extends $T_{iter}$ compared with the normal run of the application. So, we adaptively detect $T_{iter}$ of
applications while the measurement is running (line 1-7).
In line 5, we collect $\textbf{Feature}_{dect}$ after a delay of $SmpDur_{Next}$. We use the sampling sequence of $\textbf{Feature}_{dect}$
to form the curve for the period detection, so we must sample $\textbf{Feature}_{dect}$ continuously and uniformly. On the NVIDIA GPU
platform, the instantaneous power, SM utilization, and memory utilization can meet this requirement. However, the performance counters cannot
meet this requirement because its minimum sampling granularity is a CUDA kernel.
According to our experimental results, we use the composite feature of power, SM utilization, and memory utilization as $\textbf{Feature}_{dect}$, whose traces show more obvious periodicity.
In line 6, we call Algorithm \ref{alg:detection} to detect $T_{iter}$ and determine whether $T_{iter}$ is stable. If $SmpDur_{Next}>0$, we redo line 4-6 with the new $SmpDur_{Next}$. Otherwise, we think $T_{iter}$ is stable and restart the feature measurement (line 8). After a delay of $T_{iter}$, we stop measurement and collect the feature data (line 8-10).

We implement micro-intrusive online feature measurement based on Algorithm \ref{alg:measurement}. Users only need to insert the Begin and End APIs of GPOEO at the beginning and end of their source code, rather than instrument loops.



\subsection{Training Data Collection and Model Generation}
We establish energy consumption and execution time prediction models with the boosting machine learning method.
Our work aims to find the optimal SM and memory clock gears that optimize the objective function.
Therefore, these four models only need to predict the relative energy consumption and execution time relative to the NVIDIA default scheduling strategy rather than the absolute energy and time \cite{guerreiro2019modeling}.


To train the models defined in Equation \ref{equ:SM_formalize} and \ref{equ:Mem_formalize}, we define and collect four training data sets:
\begin{equation} \nonumber
	EngTr_{SM} = \{EngSM_j^i\}, \,\, TimeTr_{SM} = \{TimeSM_j^i\}
\end{equation}
\begin{equation} \nonumber
	EngTr_{Mem} = \{EngMem_j^i\}, \,\, TimeTr_{Mem} = \{TimeMem_j^i\}
\end{equation}
where $EngSM_j^i$ or $TimeSM_j^i$ represents a piece of training data
collected from application $i$ under the SM clock gear $SMgear_{j}$ and the memory clock controlled by the NVIDIA default scheduling strategy.
$EngMem_j^i$ or $TimeMem_j^i$ is collected under the memory clock gear $Memgear_{j}$ and the optimal SM clock gear.
$eng_{default}^i$ and $time_{default}^i$ are the default energy consumption and execution time under the NVIDIA default scheduling strategy.
Each input vector ${\textbf{w}_{SM}}_j^i$ or ${\textbf{w}_{Mem}}_j^i$ contains $SMgear_j$ or $Memgear_j$ and the feature vector $\textbf{Feature}^i$.
For each application, the feature vector $\textbf{Feature}^i$ is measured under the reference clock configuration.
\begin{align*}
EngSM_j^i &= \{ {\textbf{w}_{SM}}_j^i, {eng_{SM}}_j^i/eng_{default}^i\} \\
TimeSM_j^i &= \{ {\textbf{w}_{SM}}_j^i, {time_{SM}}_j^i/time_{default}^i\} \\
{\textbf{w}_{SM}}_j^i &= \{SMgear_j, \textbf{Feature}^i\} \\
EngMem_j^i &= \{ {\textbf{w}_{Mem}}_j^i, {eng_{Mem}}_j^i/eng_{default}^i\} \\
TimeMem_j^i &= \{ {\textbf{w}_{Mem}}_j^i, {time_{Mem}}_j^i/time_{default}^i\} \\
{\textbf{w}_{Mem}}_j^i &= \{Memgear_j, \textbf{Feature}^i\}
\end{align*}


\subsubsection{Feature Selection}
As mentioned in Section \ref{sssec:counter}, the high-level features, such as power, SM utilization, and memory utilization, cannot provide enough information for modeling.
So we introduce performance counter metrics as input features.

The CUPTI \cite{CUPTI} supports profiling over 1100 performance counters on NVIDIA high-end GPUs, such as RTX2080Ti and RTX3080Ti.
Inspired by Arafa’s work \cite{arafa2020verified}, we treat each Parallel Thread Execution (PTX) instruction as a basic unit of energy consumption and focus on performance counter metrics that reflect the density of different types of PTX instructions.
We list the selected metrics in Table \ref{tab:selected}. A metric with prefix \verb|sm__inst_executed_pipe_| represents a type of PTX instruction. The suffix \verb|pct_of_peak_sustained_active| means the percentage of actual mean instruction throughput to the theoretical maximum sustained instruction throughput in an activity clock cycle \cite{CUPTI}.


According to related work \cite{guerreiro2019modeling}, DRAM and L2 cache miss information may help predict energy and time.
However, profiling performance counters of DRAM and FBPA need several kernel replays which cannot be implemented online.
We design several composite metrics as alternatives, also shown in Table \ref{tab:selected}.
The \verb|L1MissPerInst| or \verb|L2MissPerInst| means the number of L1 or L2 cache miss per instruction.
The \verb|L1MissPct| or \verb|L2MissPct| means the percentage of L1 or L2 cache miss.
In conclusion, we use metrics listed in Table \ref{tab:selected} as features.

\begin{table}[t]
	\footnotesize
	\caption{Selected Features}
	\label{tab:selected}
	\begin{center}
    \rowcolors{2}{gray!25}{white}
	\begin{tabular}{p{0.18\columnwidth}p{0.7\columnwidth}}
		\toprule
		Notation      & Full name or expression \\
        \midrule
		IPCPct        & sm\_\_inst\_executed.PctSus \\
		L1MissPerInst & \begin{tabular}[c]{@{}l@{}}l1tex\_\_t\_sectors\_lookup\_miss.sum /\\ sm\_\_inst\_executed.sum\end{tabular} \\
		L1MissPct     & \begin{tabular}[c]{@{}l@{}}l1tex\_\_t\_sectors\_lookup\_miss.sum /\\ (l1tex\_\_t\_sectors\_lookup\_miss.sum +\\ l1tex\_\_t\_sectors\_lookup\_hit.sum)\end{tabular} \\
		L2MissPerInst & \begin{tabular}[c]{@{}l@{}}lts\_\_t\_sectors\_lookup\_miss.sum /\\ sm\_\_inst\_executed.sum\end{tabular} \\
		L2MissPct     & \begin{tabular}[c]{@{}l@{}}lts\_\_t\_sectors\_lookup\_miss.sum /\\ (lts\_\_t\_sectors\_lookup\_miss.sum +\\ lts\_\_t\_sectors\_lookup\_hit.sum)\end{tabular} \\
		ALUPct        & sm\_\_inst\_executed\_pipe\_alu.PctSus \\
		ADUPct        & sm\_\_inst\_executed\_pipe\_adu.PctSus \\
		FP16Pct       & sm\_\_inst\_executed\_pipe\_fp16.PctSus \\
		FMAPct        & sm\_\_inst\_executed\_pipe\_fma.PctSus \\
		FP64Pct       & sm\_\_inst\_executed\_pipe\_fp64.PctSus \\
		XUPct         & sm\_\_inst\_executed\_pipe\_xu.PctSus \\
		TNSPct        & sm\_\_inst\_executed\_pipe\_tensor.PctSus \\
		CBUPct        & sm\_\_inst\_executed\_pipe\_cbu.PctSus \\
		LSUPct        & sm\_\_inst\_executed\_pipe\_lsu.PctSus \\
		TEXPct        & sm\_\_inst\_executed\_pipe\_tex.PctSus \\
		UNIPct        & sm\_\_inst\_executed\_pipe\_uniform.PctSus \\
        \bottomrule
	\end{tabular}
	\end{center}
	\emph{Note: PctSus = sum.pct\_of\_peak\_sustained\_active} 
	
\end{table}

\subsubsection{Benchmark Selection and Training Data Collection} \label{sssec:benchmark}

We use the PyTorch Benchmarks \cite{torchbench}, which contain over 40 mini ML applications, for training.
We select the AIBench Training Component Benchmark \cite{tang2021aibench}, the benchmarking-gnns \cite{dwivedi2020benchmarking},
a ThunderSVM \cite{wenthundersvm18} workload, and a ThunderGBM \cite{8727750, wenthundergbm19, wen2019exploiting} workload as the testing set.
The benchmarking-gnns is a graph neural networks (GNN) benchmarking framework including seven datasets (CLB, CSL, SBM, TSP, TU, MLC, and SP) and nine models. 

For each application in the training set, we profile features under the reference clock configuration.
We measure the actual energy (${eng_{SM}}_j^i$) and time (${time_{SM}}_j^i$) under all available SM clock gears and the memory clock controlled by the NVIDIA default scheduling strategy.
We also measure the actual energy (${eng_{Mem}}_j^i$) and time (${time_{Mem}}_j^i$) under all available memory clock gears and the optimal SM clock gear.
Feature, energy, and time data construct four training data sets ($EngTr_{SM}$, $TimeTr_{SM}$, $EngTr_{Mem}$, and $TimeTr_{Mem}$).
We conduct each measurement ten times and take the average to reduce the training data noise.

\subsubsection{Prediction model} \label{sssec:training}

We construct energy and time prediction models with the XGBoost \cite{chen2016xgboost}, a widely used supervised machine learning model improved from the gradient tree boosting algorithm.
Combined with our scenario, we apply the XGBoost to build four prediction models shown in Equation \ref{equ:SM_formalize} and \ref{equ:Mem_formalize}.
The model construction processes are similar, so we take the energy consumption model $EngMdl_{SM}$ as an example to introduce these processes.
For $EngMdl_{SM}$, the XGBoost weights and sums several regression trees and can be defined as follows:
\begin{equation} \nonumber
\widehat{{eng_{SM}}_j^i} = \sum_{k=1}^Kf_k({\textbf{w}_{SM}}_j^i), f_k \in \mathcal{F}
\end{equation}
where $\widehat{{eng_{SM}}_j^i}$ is the predicted energy consumption of the input vector ${\textbf{w}_{SM}}_j^i$, which contains the features of application $i$ and the SM gear $j$.
$f_k$ represents the $k$th regression tree, and $K$ is the number of regression trees.
We use the training datasets $EngTr_{SM}$ to train the model $EngMdl_{SM}$. During training, we use the grid search method to tune the hyper-parameters such as the minimum loss reduction, the maximum depth of a tree, the minimum sum of instance weight, and the maximum number of nodes to be added.

\subsubsection{Online Local Search}
To fix minor errors caused by multi-objective predicted models, we perform a local search around the predicted optimal clock configuration. We use online detected period and energy data measured during one period to evaluate different clock frequencies and find the actual optimal clock configuration.
We first conduct a local search for the memory clock because wrong memory clock frequencies (too low) can lead to severe slowdowns. Based on the optimal memory clock frequency, we conduct another local search for the SM clock.
According to our experimental results and Schwarzrock’s work \cite{schwarzrock2020runtime}, the relationship between energy optimization objectives and clock gears is generally a convex function.
Based on this observation, we use the golden-section search method \cite{walser2001golden} to accelerate our local search.
We first find a gear with a worse objective value on each side of the predicted optimal gear to determine the search interval.
Then we follow the classic golden section-search process \cite{walser2001golden} to attempt different clock gears.
Due to possible errors of measured energy and periods, we fit the attempted points to obtain a convex function.
Then we use the convex function to determine the optimal clock gear.

\begin{figure*}[t]
	\centering
	\includegraphics[width=7in]{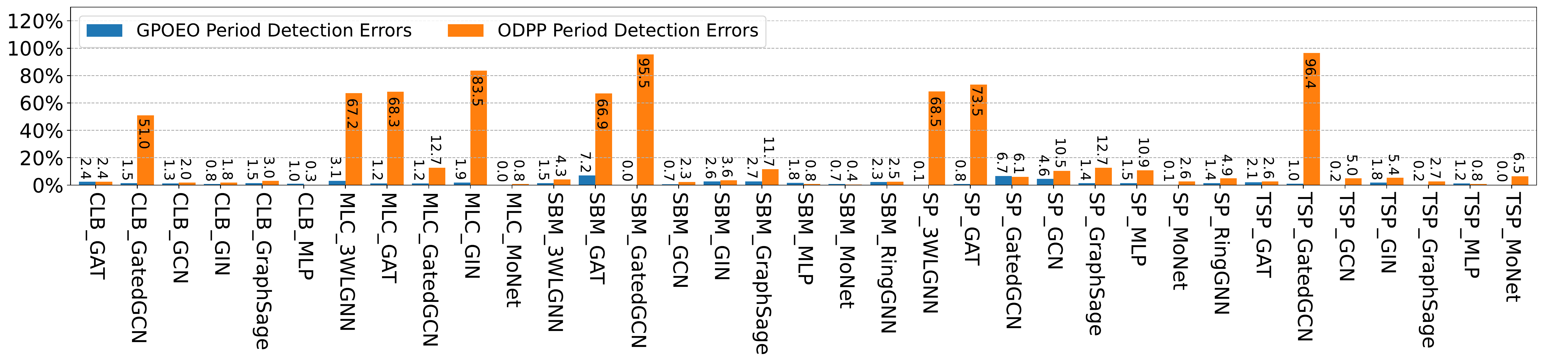}
	\caption{The period detection errors of GPOEO and ODPP. GPOEO gives lower errors compared to ODPP. }
	\label{fig:PeriodErr}
\end{figure*}

\subsubsection{Optimize aperiodic applications}
`
The power and utilization traces of some applications are aperiodic, such as applications in CSL and TUs datasets, the ThunderSVM \cite{wenthundersvm18} workload, and the ThunderGBM \cite{8727750, wenthundergbm19, wen2019exploiting} workload.
We can not evaluate their execution time or energy consumption with data measured in one period.
In this case, we fix the measurement time interval and the use measured mean number of instructions executed per second (IPS) and power to evaluate the execution time and energy consumption.
If the number of instructions in the program is $Inst_{sum}$, the execution time can be calculated as $time = Inst_{sum}/IPS$, and the energy consumption can be calculated as $energy = power * time = Inst_{sum} * power / IPS$.
Then we can evaluate different clock frequency configurations.

Based on these analyses, we apply GPOEO to aperiodic applications. We first measure performance counter metrics in the fixed time interval and
predict the optimal clock frequency configuration. In the online local search, we measure and calculate $time_{default}$ and $energy_{default}$
under the NVIDIA default scheduling strategy as the baseline. Then, we follow the golden-section search strategy and collect $time_{i}$ and
$energy_{i}$ under different clock frequency configurations near the predicted optimal clock frequency configuration.
Finally, we compare these $time_{i}$ and $energy_{i}$ with the baseline to find the optimal clock frequency configuration.

\section{Evaluation} \label{sec:evaluation}

We evaluate our GPOEO system in this section. We first introduce the experimental step. Then, we evaluate the accuracy of the period detection algorithm and energy-performance prediction models. Finally, we analyze the online energy-saving results on various ML benchmarks.

\subsection{Experimental Setup}

\subsubsection{Hardware and Software Platforms}
Our experimental platform is a GPU server equipped with one NVIDIA RTX3080Ti GPU, one AMD 5950X CPU, and 64 GB memory. 
The software environment is NVIDIA Driver 470.57 and CUDA 11.3.
RTX3080Ti supports continuously adjustable SM clock frequency, from 210 MHz to 2,025 MHz, and the step is 15 MHz.
We find that some higher frequencies are not practical or stable.
Under lower frequencies, applications cannot save energy while suffering severe slowdowns.
Therefore, we only consider the frequencies in the middle part, which can run stably and may improve energy efficiency.
We treat each SM clock frequency as an SM clock gear, from $SMgear_{16}$ = 450 MHz to $SMgear_{114}$ = 1,920 MHz. 
RTX3080Ti supports 5 global memory clock frequencies: $Memgear_{0}$ = 450 MHz, $Memgear_{1}$ = 810 MHz, $Memgear_{2}$ = 5,001 MHz, $Memgear_{3}$ = 9,251 MHz, and $Memgear_{4}$ = 9,501 MHz.
We select $SMgear_{106}$ = 1,800 MHz and $Memgear_{3}$ = 9,251 MHz as reference frequencies, which are used in performance counter profiling.




\subsubsection{Benchmarks}

We train our machine learning models using data collected from PyTorch Benchmarks \cite{torchbench}. We then test the
trained models on AIBench \cite{tang2021aibench} and benchmarking-gnns \cite{dwivedi2020benchmarking}. In addition to
deep neural networks, we also use two classical ML workloads, the ThunderSVM \cite{wenthundersvm18} and ThunderGBM \cite{8727750, wenthundergbm19, wen2019exploiting}, in our evaluation.

\begin{figure}[t]
	\centering
	\includegraphics[width=3.5in]{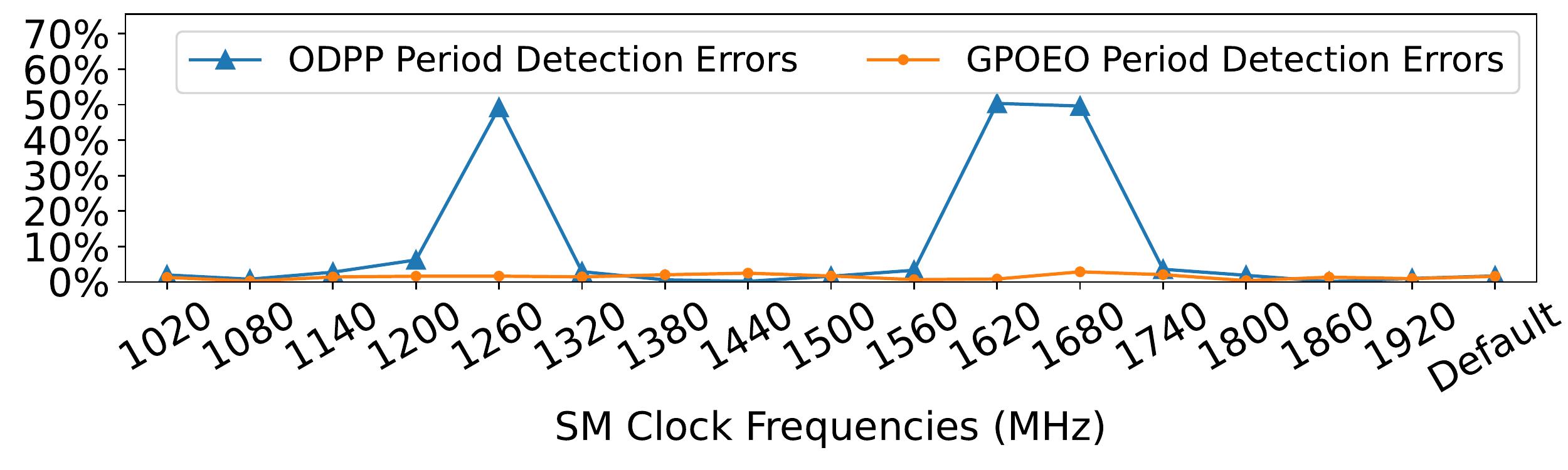}
	\caption{Period detection errors of ODPP and GPOEO on the CLB\_GAT application under different SM clock frequencies}
	\label{fig:Sensitivity-COLLAB_GAT}
\end{figure}

\begin{figure}[t]
	\centering
	\includegraphics[width=3.5in]{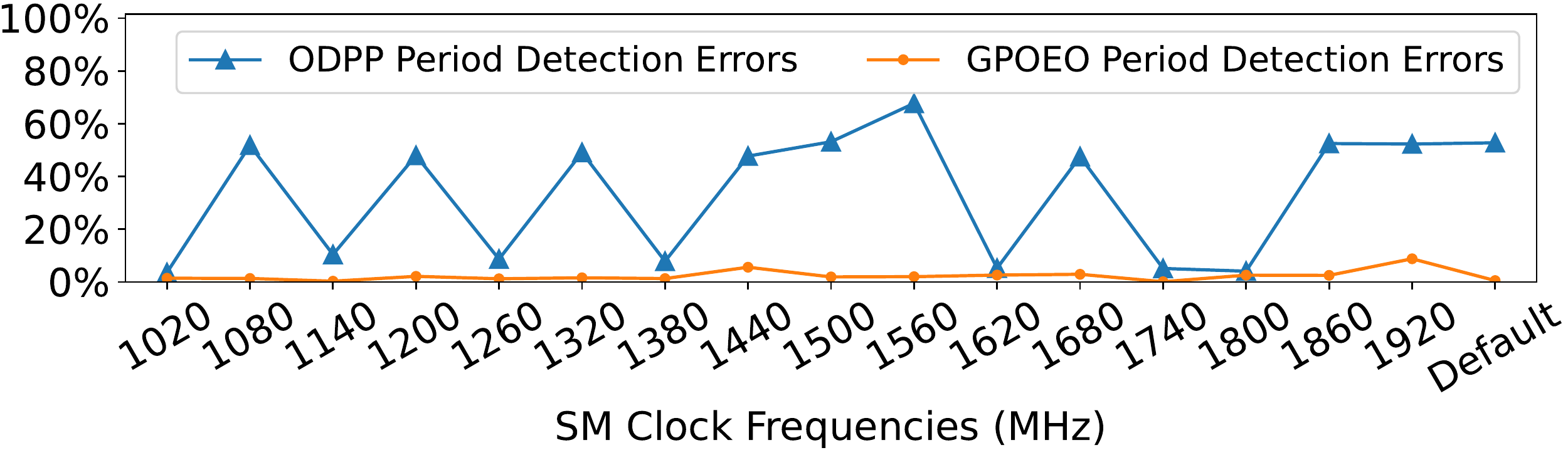}
	\caption{Period detection errors of ODPP and GPOEO on the SBM\_3WLGNN application under different SM clock frequencies}
	\label{fig:Sensitivity-SBMs_3WLGNN}
\end{figure}

\begin{figure}[t]
	\centering
	\includegraphics[width=3.5in]{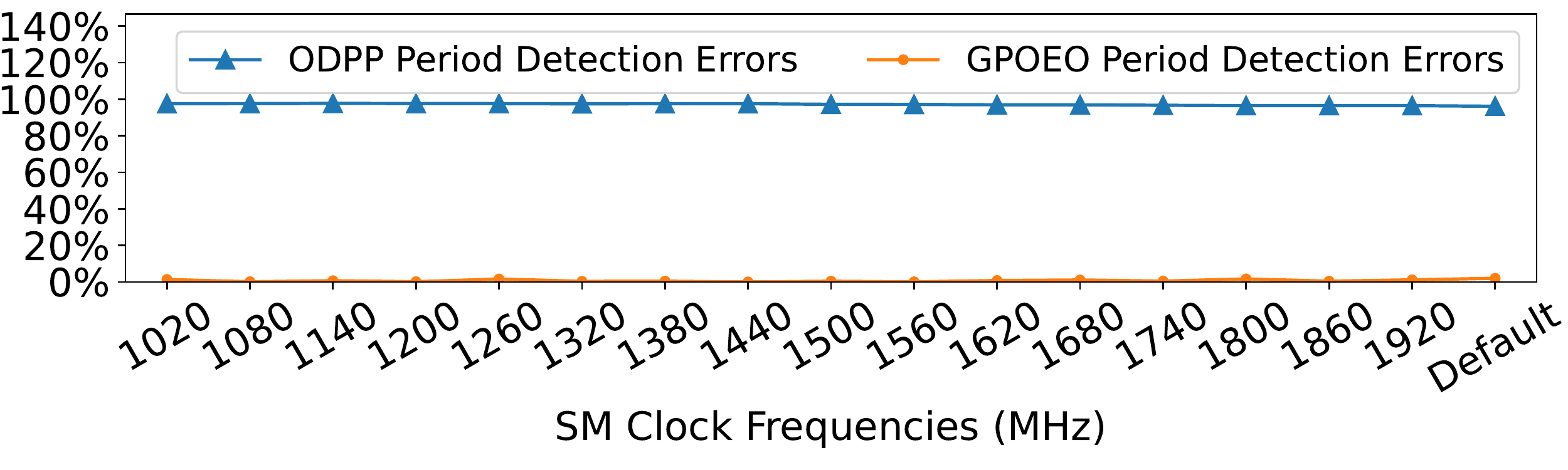}
	\caption{Period detection errors of ODPP and GPOEO on the TSP\_GatedGCN application under different SM clock frequencies}
	\label{fig:Sensitivity-TSP_GatedGCN}
\end{figure}

\begin{figure}[t] \centering
	\subfigure[Energy consumption prediction errors] {
		\includegraphics[width=3.5in]{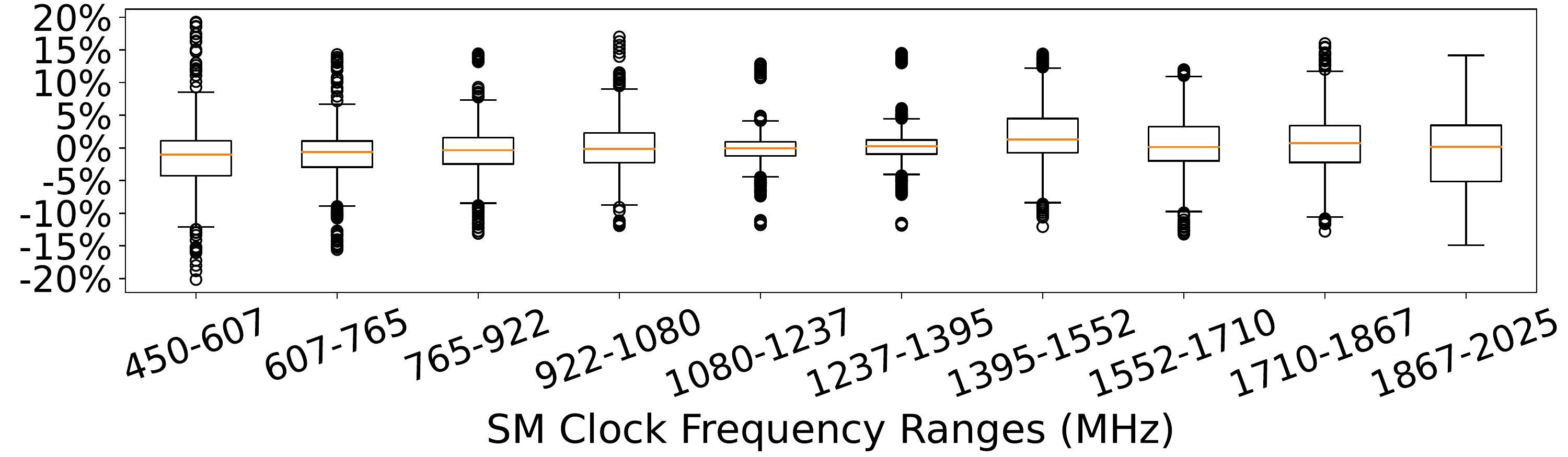}
	}
	\subfigure[Execution time prediction errors] {
		\includegraphics[width=3.5in]{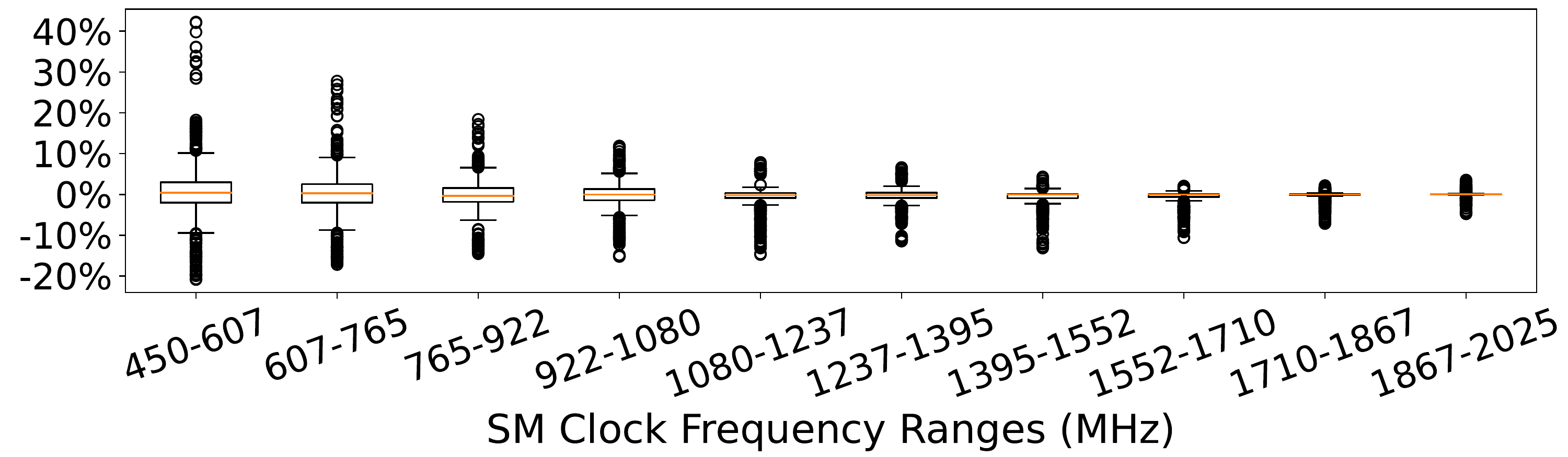}
	}
	\caption{Prediction errors by varying the SM clock (grouped by different SM clock ranges)}
	\label{fig:ErrByClock}
\end{figure}

\begin{figure}[!h] \centering
	\subfigure[Energy consumption prediction errors] {
		\includegraphics[width=3.5in]{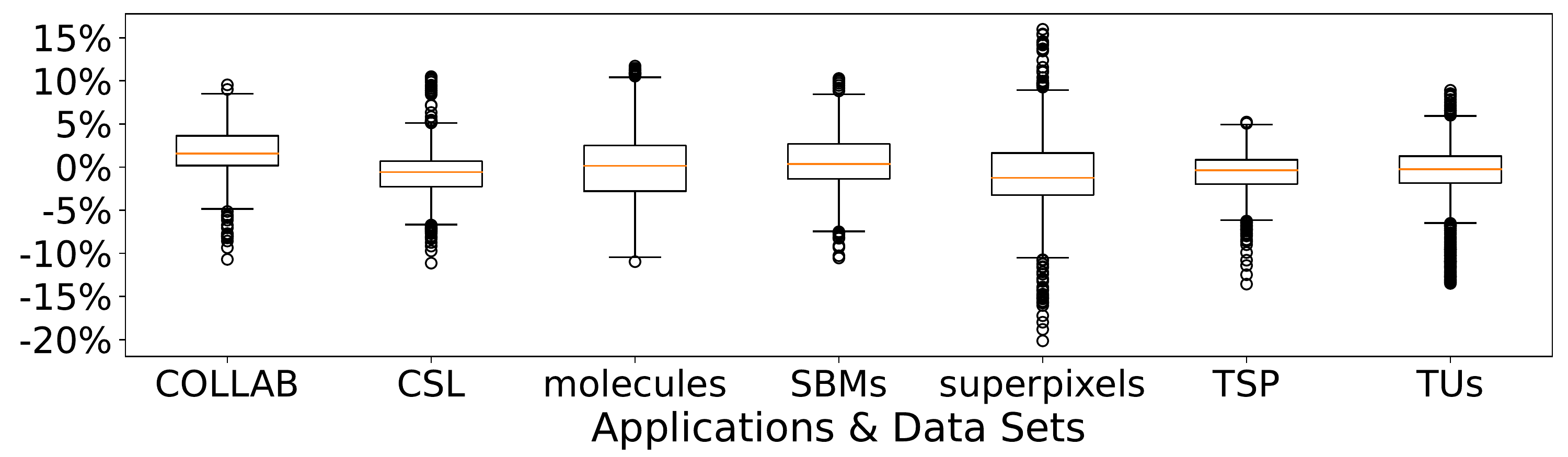}
	}
	\subfigure[Execution time prediction errors] {
		\includegraphics[width=3.5in]{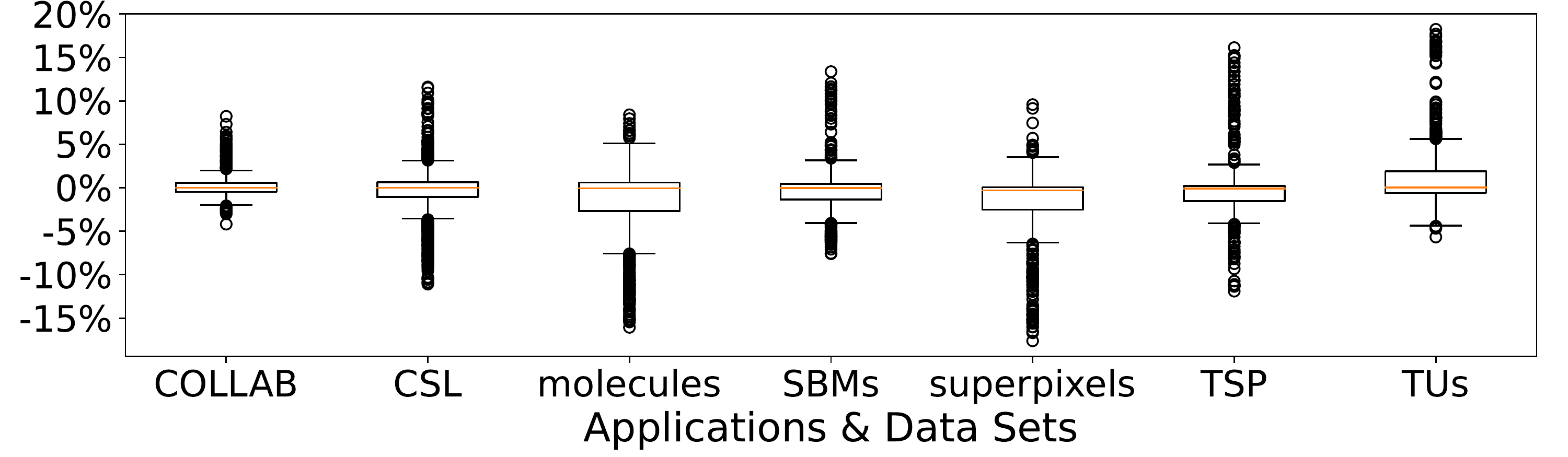}
	}
	\caption{Prediction errors by varying the SM clock (grouped by different datasets)}
	\label{fig:ErrByApp}
\end{figure}

%
%
%

\subsection{Accuracy and Sensitivity of Period Detection} \label{ssec:period_accuracy}

We analyze our period detection algorithm on 34 different ML applications. Figure \ref{fig:PeriodErr} shows the period
detection errors of GPOEO and ODPP \cite{ODPP} using the NVIDIA default scheduling strategy. GPOEO is far more accurate
than ODPP. The mean period error of GPOEO is 1.72\%, while the mean error of ODPP is 23.16\%. The maximum error of
GPOEO is 7.2\%, and 32 errors are within 5\%. For ODPP, nine errors are above 50\%, nine errors are among 5-13\%, and
sixteen errors are within 5\%. GPOEO is more accurate than ODPP on 29 applications. For the other three applications,
the accuracy differences are within 1\%.


\begin{figure}[t] \centering
	\subfigure[Energy consumption prediction errors] {
		\includegraphics[width=3.5in]{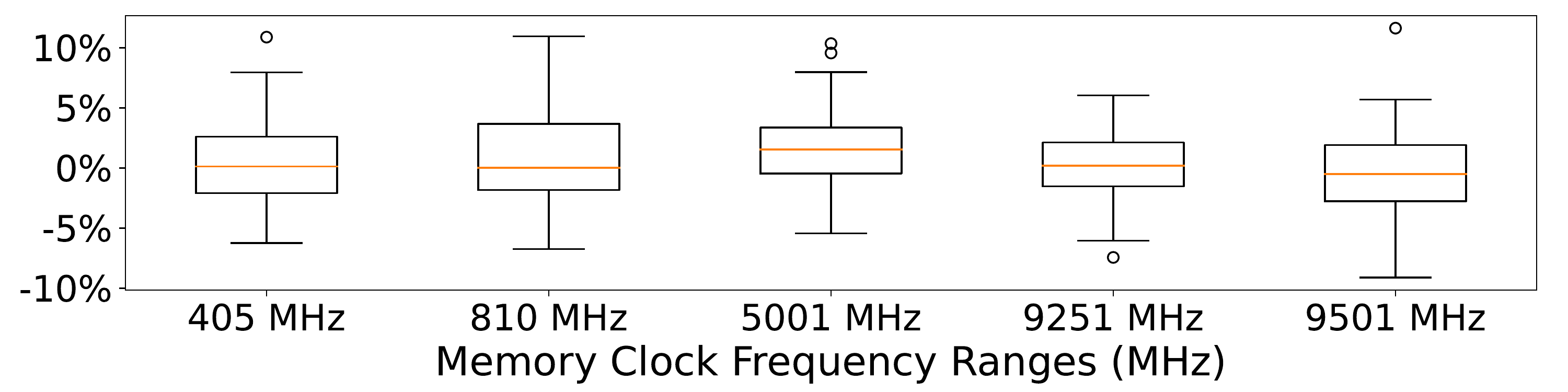}
	}
	\subfigure[Execution time prediction errors] {
		\includegraphics[width=3.5in]{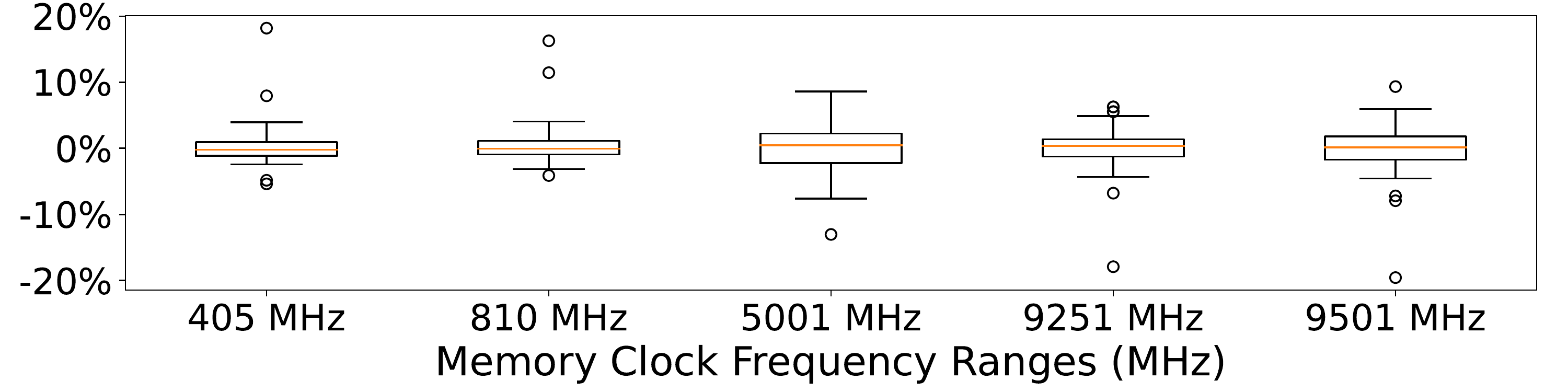}
	}
	\caption{Prediction errors by varying the memory clock (grouped by different memory clock frequencies)}
	\label{fig:ErrByClock_Mem}
\end{figure}

\begin{figure}[t] \centering
	\subfigure[Energy consumption prediction errors] {
		\includegraphics[width=3.5in]{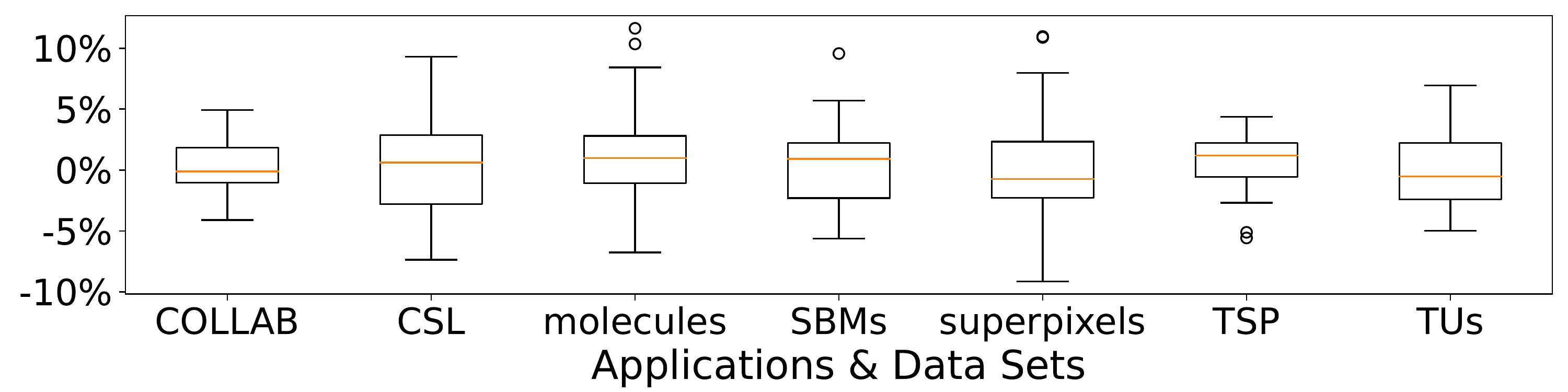}
	}
	\subfigure[Execution time prediction errors] {
		\includegraphics[width=3.5in]{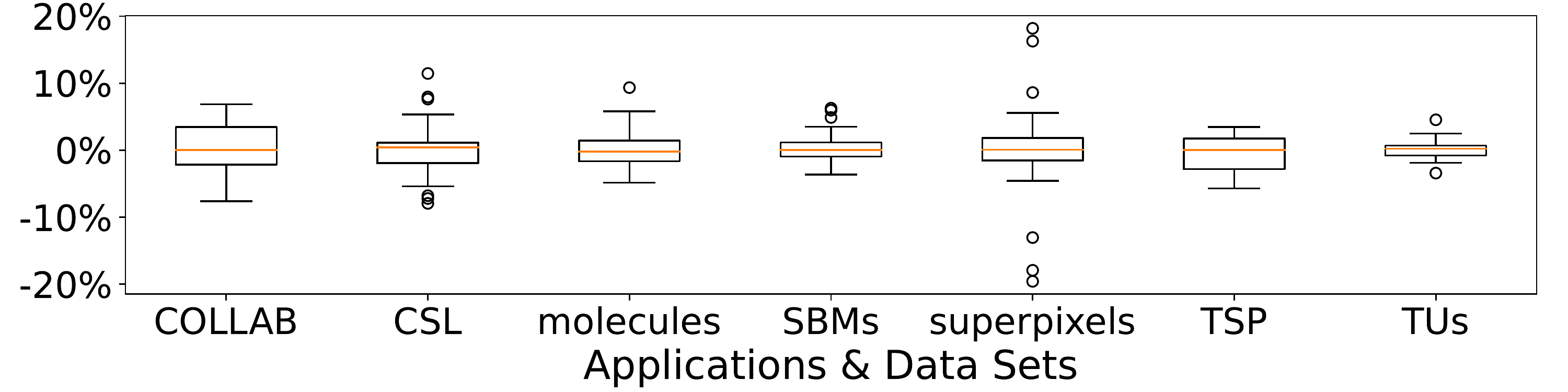}
	}
	\caption{Prediction errors by varying the memory clock (grouped by different datasets)}
	\label{fig:ErrByApp_Mem}
\end{figure}

%
%
%

Figure \ref{fig:Sensitivity-COLLAB_GAT}, \ref{fig:Sensitivity-SBMs_3WLGNN}, and \ref{fig:Sensitivity-TSP_GatedGCN} show the periods detection
errors of GPOEO and ODPP under varying SM clock frequencies and fixed memory clock frequency.
GPOEO exhibits stable low errors (within 5\% for most cases). This phenomenon indicates that GPOEO is sensitive to clock frequency changes.
However, ODPP shows poor stability. On CLB\_GAT and SBM\_3WLGNN, ODPP gains huge errors under three and ten SM clock
frequencies, respectively. This instability causes inaccurate prediction models, which ODPP builds online.
For TSP\_GatedGCN, ODPP gains huge errors (nearly 100\%) under all frequencies because it is disturbed by high-frequency noise, only detects the local short period, and ignores the real global period.



\subsection{Accuracy of Energy and Execution Time Prediction}

In this section, we evaluate four prediction models.
We measure selected features (Table \ref{tab:selected}) during one iteration of each ML application under the reference SM and memory clock frequency (1800 MHz and 9251MHz).
Then we use these features and different SM and memory clock frequencies as input of our multi-objective models to predict the relative energy consumption and execution time relative to the NVIDIA default scheduling strategy.
Note that for the SM clock models $EngMdl_{SM}$ and $TimeMdl_{SM}$, we let the NVIDIA default scheduling strategy control the memory clock and then predict and select the optimal SM clock frequency.
For the memory models $EngMdl_{Mem}$ and $TimeMdl_{Mem}$, we assume that the SM clock is already set to the optimal frequency and then predict and select the optimal memory clock frequency.

For the SM clock models, we collect 11,660 prediction values in total (2 objectives, 55 applications, and 106 SM clock frequencies).
The mean prediction errors of energy consumption and execution time are 3.05\% and 2.09\%, respectively.
Next, we analyze the distributions of errors sorted by SM clock frequency ranges and application datasets.
Figure \ref{fig:ErrByClock} shows the prediction errors of energy consumption and execution time
separately grouped by different SM clock frequency ranges.
Each clock frequency range contains 550 or 605 prediction values depending on the number of energy gears included in the clock range (ten or eleven gears).
For all clock ranges, most energy and time prediction errors are less than 5\%.
Most scattered errors are within 20\%. This phenomenon proves that our energy prediction model and execution time prediction model are accurate for different SM clock frequencies.
Figure \ref{fig:ErrByApp} shows the prediction errors grouped by different application datasets.
Each group contains 636 to 954 predicted values depending on the number of applications included in the datasets (six to nine applications).
For all application datasets, most energy and execution time prediction errors are less than 5\%.
This phenomenon demonstrates that our SM clock models are applicable to all these applications.

For the memory clock models, we collect 550 prediction values in total (2 objectives, 55 applications, and 5 memory clock frequencies).
The mean prediction errors of energy consumption and execution time are 2.72\% and 2.31\%, respectively.
Figure \ref{fig:ErrByClock_Mem} shows the prediction errors of energy consumption and execution time separately grouped by different memory clock frequencies.
For all clock frequencies, most energy and time prediction errors are less than 5\%.
Figure \ref{fig:ErrByApp_Mem} shows the prediction errors grouped by different application datasets.
For all application datasets, most prediction errors are also within 5\%.
These phenomena demonstrate that our memory clock models are applicable to all memory frequencies and all these applications.

\begin{figure*}[t]
	\centering
	\includegraphics[width=7in]{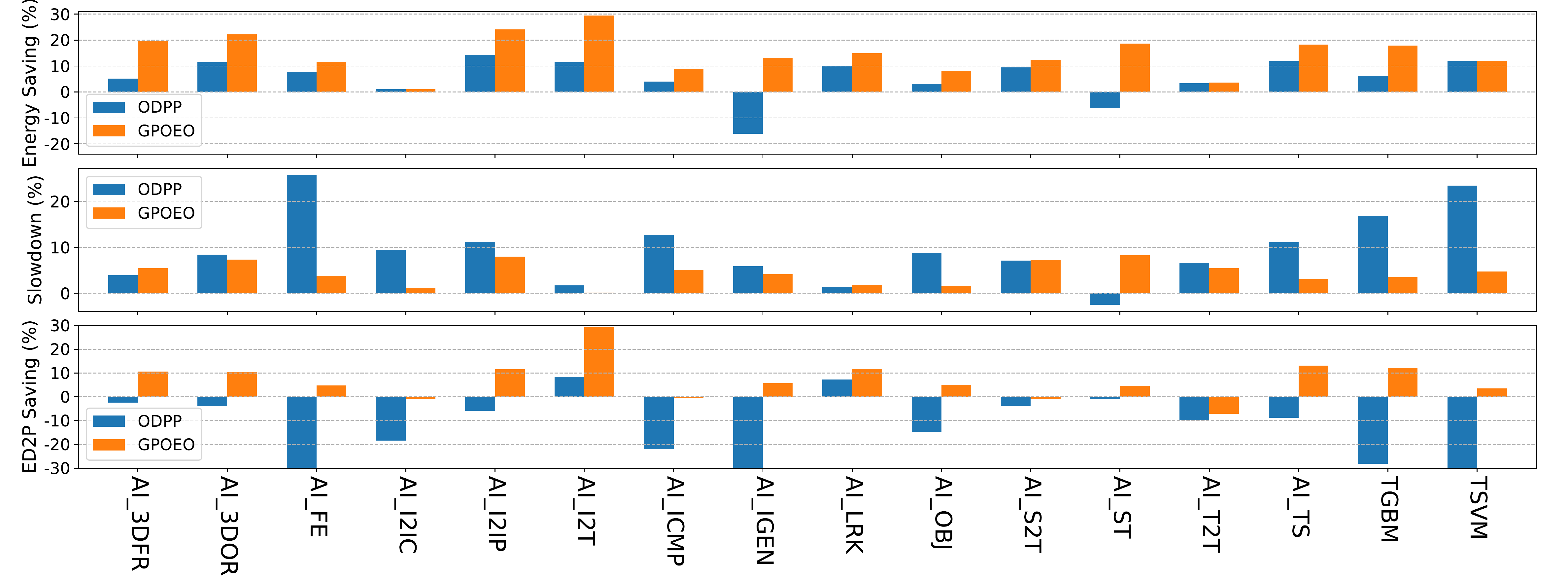}
	\caption{Energy savings, slowdown, and ED2P saving compared to the NVIDIA default scheduling strategy on AIBench and traditional ML applications}
	\label{fig:Result_AIBench}
\end{figure*}

\begin{table*}[!h]
	\footnotesize
	\caption{Online optimization process for SM and memory clock on AIBench} 
	\label{tab:OptProcess}
	\begin{center}
		\rowcolors{2}{gray!25}{white}
		\begin{tabular}{p{0.38\columnwidth}lllllllllllllllll}
			\toprule
			& 3DFR & 3DOR & FE & I2IC & I2IP & I2T & ICMP & IGEN & LRK & OBJ & S2T & ST & T2T & TS \\  
            \midrule
			Oracle SM Gear                   & 104   & 102   & 109     & 104  & 91   & 94   & 107  & 71   & 88  & 100  & 108 & 39 & 100  & 107 \\ 
			Prediction Error (SM gear)   &  -11  & 5   &  -13   & 2  & 3  & -2  & -13  & 22   & 24   & -4 & -15 & 11 & 5 & -5 \\ 
			Search Error (SM gear) & -2   & -3   & -8     & 1   & -2   & 0    & -2    & -3    & -2   & 0   & -7  & 4 & -2 & -1 \\ 
			\# of Search Steps (SM)	& 5    & 4   & 4    & 3   & 4    & 3   & 6   & 8    & 9   & 4  & 5  & 5  & 4   & 4 \\ 
			Oracle Mem clock (MHz) & 9501 & 9251 & 9501 & 9501 & 9251 & 9251 & 9251 & 405 & 5001 & 9251 & 9251 & 810 & 9251 & 9501 \\ 
			Predicted Mem clock &  9501 & 9501 & 9251 & 9501 & 9501 & 9501 & 9501 & 405 & 5001 & 9251 & 9501 & 405 & 9251 & 9501 \\ 
			Searched Mem clock & 9501 & 9501 & 9501 & 9501 & 9501 & 9251 & 9251 & 405 & 5001 & 9251 & 9251 & 810 & 9251 & 9501 \\ 
			\# of Search Steps (Mem)	& 2    & 2   & 3    & 2   & 2    & 2   & 2   & 3    & 3   & 3  & 2  & 3  & 3   & 2 \\
            \bottomrule
		\end{tabular}
		
	\end{center}
\end{table*}

\subsection{Results of Online Optimization} \label{ssec:ED2P}

Our multi-objective models predict the energy consumption and execution time of one iteration under different SM and memory clock gears. Users can use diverse energy efficiency objectives to select the predicted optimal energy gears and guide local search to find the actual optimal energy gears.
In this experiment, we set the objective function to minimize the energy consumption within the slowdown constraint of 5\%.
For each ML training application, we use the energy consumption and execution time under the NVIDIA default scheduling strategy
as the baseline. Then we run these applications with our GPOEO system to optimize the energy consumption online. We also
implement the ODPP \cite{ODPP} as a comparison.


\subsubsection{Medium benchmark suite} Figure \ref{fig:Result_AIBench} shows the online optimization results of the
AIBench, the ThunderSVM workload, and the ThunderGBM workload. We use energy saving, slowdown (execution time
increase), and $ED^2P$ ($ED^2P=Energy \times Time^2$) saving to evaluate GPOEO and ODPP \cite{ODPP}. GPOEO achieves an
average energy saving of 14.7\% and an average $ED^2P$ saving of 6.8\%, with an average execution time increase of
4.6\%. Especially, GPOEO achieves significant energy saving ($\geq20\%$) on four out of sixteen applications. On
AI\_I2T, GPOEO achieves 29.5\% energy saving and 0.1\% slowdown. Table \ref{tab:OptProcess} shows the online
optimization process for SM and memory clocks. On AI\_I2T, the error of our multi-objective prediction models is only
-2 SM gears. This error is eliminated by the online local search within five steps so that the vast energy saving can
be achieved. For AI\_3DFR, AI\_3DOR, AI\_I2IP, and AI\_TS, GPOEO obtains considerable energy savings (18-24\%) with
reasonable slowdowns (less than 8\%). The online optimization processes of these applications are similar to AI\_I2T.
The prediction errors of SM gear are relatively small (within 11 SM gears), and the online local search reduces these
errors in three to five steps.

AI\_FE, AI\_ICMP, AI\_IGEN, AI\_LRK, AI\_OBJ, AI\_S2T, and AI\_ST have medium energy-saving opportunities (10-21\%) with the slowdown constraint of 5\%.
GPOEO gains medium energy savings (8-14\%) with reasonable slowdowns (2-7\%) on these seven applications.
Two reasons cause GPOEO does not make full use of energy-saving potentials. For AI\_FE and AI\_S2T, the period detection and feature measurement get inaccurate data due to some abnormal iterations. Therefore prediction errors are significant and local search can not eliminate these errors.
For the convenience of experiments, we just train each machine learning model for a few iterations, and only iterations after optimization processes can enjoy the optimal SM gear. However, in real-life scenarios, the model will be trained on many more iterations, for which our approach can give higher energy saving.
GPOEO does not perform well on AI\_I2IC and AI\_T2T.
These two applications are computation-intensive with little energy-saving opportunities (1\% and 4\%).
For non-periodical applications TGBM and TSVM, GPOEO also achieves good results, saving 17.9\% and 12.0\% energy with slowdowns of 3.5\% and 4.7\%, respectively.
As for the memory clock, the oracle frequencies of thirteen applications are 9,251 MHz or 9,501 MHz. Our method finds oracle frequencies on eleven applications and finds nearby 9,501 MHz on the other two applications within two or three search steps.
In fact, the energy consumption and execution time are similar under these two memory clock frequencies.
GPOEO also finds low oracle memory frequencies on AI\_IGEN, AI\_LRK, and AI\_ST within three steps.
Our method meets the performance loss constraint of 5\% on nine applications, while ODPP only achieves the same goal on four applications.
For GPOEO, only iterations after optimization processes are guaranteed to meet the performance loss constraint, and the previous iterations may result in a violation of the performance loss constraint.

\begin{figure*}[h]
	\centering
	\includegraphics[width=7in]{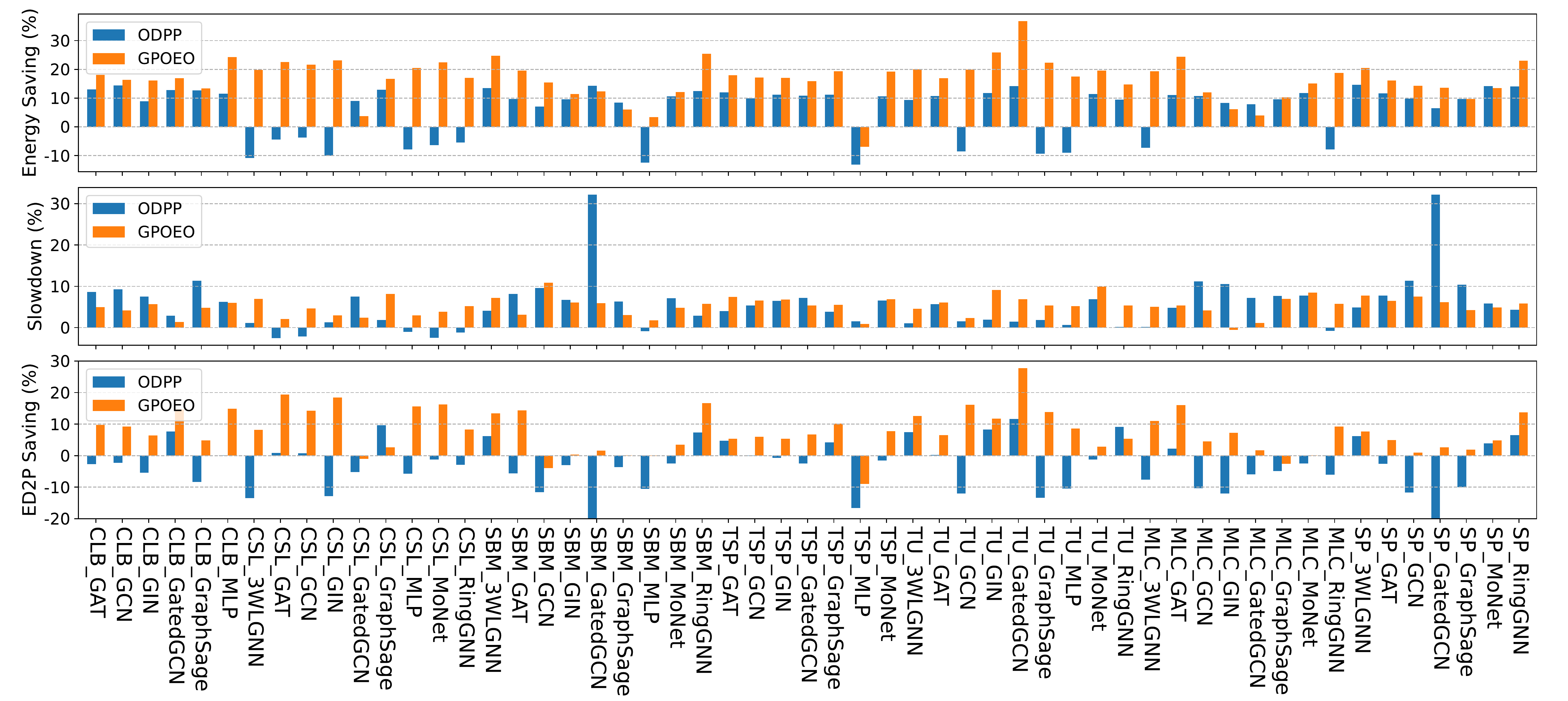}
	\caption{Energy saving, slowdown, and ED2P saving compared to the NVIDIA default scheduling strategy on benchmarking-gnns applications}
	\label{fig:Result_GNN}
\end{figure*}

Compared to ODPP, our approach can improve the energy saving by 9.2\%, improve the $ED^2P$ saving by 20.4\%, and reduce the slowdown by 5.0\% on average.
GPOEO performs better than ODPP on all sixteen applications.
Especially, our approach gains both larger energy savings and lighter slowdowns on eight applications.
For AI\_3DFR, AI\_GEN, AI\_LRK, and AI\_S2T, GPOEO achieves more considerable energy savings with similar slowdowns.
Our method gains similar energy savings with lighter slowdowns on AI\_I2IC and TSVM.
ODPP builds two piecewise linear models online to predict energy and time.
The accuracy of these two models is highly dependent on the period detection accuracy.
Poor period detection accuracy of ODPP causes significant prediction errors, so ODPP shows heavier slowdowns and less energy
saving. On non-periodical applications TGBM and TSVM, ODPP shows much worse results than GPOEO.
This phenomenon demonstrates that ODPP cannot handle non-periodical applications, while GPOEO can.



\subsubsection{Large benchmark suite}



We also use the benchmarking-gnns suite to evaluate our method and ODPP \cite{ODPP} extensively.
Figure \ref{fig:Result_GNN} shows the online optimization results of 55 applications in the benchmark-gnns.
GPOEO achieves an energy saving of 16.6\% and an $ED^2P$ saving of 7.8\%, with an execution time increase of 5.2\% on average.
In comparison, ODPP gains an energy saving of 6.1\% and an $ED^2P$ saving of -4.5\%, with an execution time increase of 5.6\%
on average. GPOEO saves $ED^2P$ on 50 applications, while ODPP only gains $ED^2P$ saving on 19 applications.
Our method meets the performance loss constraint of 5\% on 22 applications.
ODPP achieves the same goal on 27 applications because it does not profile performance counters and has a lower measurement
overhead. With the help of performance counter features, GPOEO saves more energy than ODPP on 48 applications.
Applications in CSL and TU datasets are non-periodical.
ODPP can not handle these applications and only saves 0.73\% of energy with a slowdown of 1.32\% on average.
On the contrary, GPOEO gains an energy saving of 20.0\% with a slowdown of 5.2\% on average.
These phenomena show that GPOEO has better performance and broader applicability than ODPP.


\begin{figure}[t]
	\centering
	\includegraphics[width=3.5in]{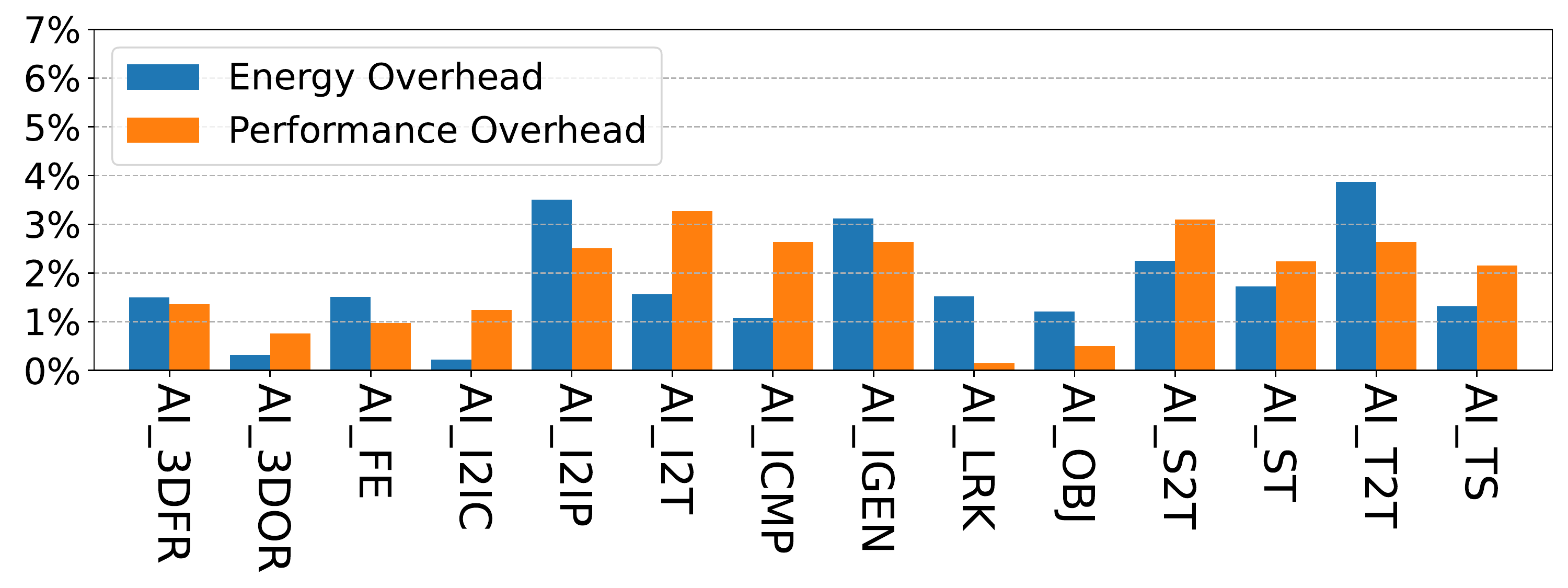}
	\caption{Energy and time overhead of GPOEO on AIBench}
	\label{fig:Overhead}
\end{figure}

\subsection{Overhead Evaluation}

To evaluate the overhead of the GPOEO system, we run AIBench applications with the entire GPOEO system except adjusting SM and memory clocks.
We also manually set the number of local search steps shown in Table \ref{tab:OptProcess}. 
As shown in Figure \ref{fig:Overhead}, all energy and time overheads are within 4\%.

\section{Related Work} \label{sec:Related_Work}
There is a growing interest in optimizing machine learning workloads for performance and energy efficiency.
Many works concentrate on CPU energy optimization
\cite{endrei2019statistical,ramesh2019understanding,hao2020fine,gholkar2019uncore,wang2021online,schwarzrock2020runtime,petoumenos2015power}.
Some of these prior works \cite{endrei2019statistical,ramesh2019understanding,hao2020fine} use offline methods to
model energy and performance, while others \cite{gholkar2019uncore,wang2021online,schwarzrock2020runtime} design
runtime to adjust the CPU clock or enforce power capping \cite{petoumenos2015power}. These methods rely heavily on
run-time CPU-specific performance counter profiling and are not transferable to GPUs, but the prediction models
\cite{endrei2018energy, hao2020fine, wang2021online} they have used may also apply to GPUs.

Profiling workloads using GPU performance counter during program execution time can be expensive. For this reason, most
GPU energy optimization works have to rely on offline profiling
information \cite{fan2019predictable,guerreiro2019modeling,guerreiro2019ptx,arafa2020verified}. Most offline works
\cite{fan2019predictable,guerreiro2019modeling,guerreiro2019ptx} focus on modeling and predicting energy and
performance under different hardware configurations. Other works study GPU energy and performance from the view of
assembly instructions \cite{guerreiro2019ptx,arafa2020verified}.
These works require tedious offline profiling and analysis for each new application or even different inputs. In
contrast, our work can automatically optimize new applications online once the prediction models are established and
well trained. Although these offline jobs are inconvenient to use, the energy features
\cite{fan2019predictable,guerreiro2019modeling} they have used are instructive for GPU online energy consumption
optimization.

A few studies \cite{majumdar2017dynamic,ODPP} realize GPU online energy consumption optimization. Majumdar \emph{et
al.} \cite{majumdar2017dynamic} exploit low-overhead fine-grained profiling functions supported by the APU platform. It
obtains the stream of energy, execution time, throughput, and performance counter metrics at kernel granularity without
worrying about time and energy overhead. Therefore, it can catch energy-saving chances at kernel granularity and save
considerable energy with low overhead. Majumdar's work cannot be transplanted to the NVIDIA GPU platform, considering
the significant amount of code instrumentation (around each kernel) and the high overhead of continuously profiling.
Our approach targets the NVIDIA GPU platform with high runtime profiling overhead (with a slowdown of $>8\%$ and energy
overhead of  $>10\%$). We profile performance counter metrics in only one iteration period to minimize overhead. ODPP
\cite{ODPP} proposes an online dynamic power-performance management framework for GPUs using coarse-grained features
which cannot provide enough information to model energy consumption and performance accurately, as mentioned in Section
\ref{sssec:counter}. Besides, the period detection algorithm of ODPP is unstable and error-prone. Due to the two
weaknesses, ODPP exhibits poor energy-saving results and can give severe slowdown in our evaluation. Our approach
exploits hardware performance counter metrics and utilizes a robust period detection algorithm to tackle these two
weaknesses, leading to better performance.

Energy and performance optimization built upon machine learning techniques have been demonstrated to be promising in
various application domains
\cite{wang2009mapping,grewe2013portable,wang2014automatic,cummins2017end,wen2014smart,taylor2017adaptive,wang2018machine,
fang2020parallel,zhang2018auto,stamoulis2018hyperpower,wang2020energy, marco2020optimizing,
ren2020camel,nabavinejad2021batchsizer,brownlee2021exploring, zhang2020optimizing, qiu2021optimizing, ye2020deep}. Some
works \cite{stamoulis2018hyperpower,wang2020energy,nabavinejad2021batchsizer,brownlee2021exploring,marco2020optimizing,
qiu2021optimizing} tune hyperparameters, model structure configurations, or data structures to speed up ML applications
or save energy consumption. Our method can be used together with these application-level techniques to reduce energy
consumption further. CAMEL \cite{ren2020camel} optimizes the energy of web applications on mobile platforms. It uses
the frame rate as the performance index, which is only suitable for web applications. Many studies
\cite{cummins2017end,taylor2017adaptive,wang2009mapping,grewe2013portable,wen2014smart,zhang2018auto,
zhang2020optimizing,fang2020parallel,wang2018machine} concentrate on accelerating applications on heterogeneous
platforms. Combining these studies with our method may enable acceleration while saving energy. Based on the graph
neural network, Ye \emph{et al.} \cite{ye2020deep} propose a novel program analysis and modeling method to provide
comprehensive insight into the program. This method may be helpful to build more accurate energy and performance
prediction models.

\section{Conclusion}
We have presented GPOEO, a new online GPU energy optimization framework for iterative machine learning (ML) workloads.
GPOEO detects iterative periods, measures performance counter features, and predicts the optimal energy configurations
automatically. We evaluate GPOEO on 71 ML applications running on an NVIDIA RTX3080Ti GPU. GPOEO achieves a mean energy
saving of 16.2\% at a modest cost of 5.1\% performance loss compared to the NVIDIA default scheduling strategy.

\ifCLASSOPTIONcompsoc
\section*{Acknowledgments}
\else
\section*{Acknowledgment}
\fi This work was supported in part by the National Key Research and Development Program of China (2020YFB1406902), the
Key-Area Research and Development Program of Guangdong Province (2020B0101360001), the Shenzhen Science and Technology
Research and Development Foundation (JCYJ20190806143418198), the National Natural Science Foundation of China (NSFC)
(61872110 and 61872294), the Fundamental Research Funds for the Central Universities (Grant No.  HIT.OCEF.2021007), the
Peng Cheng Laboratory Project (PCL2021A02), and the CCF-Huawei fund (CCF-HuaweiHP2021002). Professor Weizhe Zhang is
the corresponding author.


\bibliographystyle{IEEEtran}
\bibliography{reference.bib}

\end{document}